\documentclass[fleqn,usenatbib]{mnras}
\usepackage[T1]{fontenc}

\DeclareRobustCommand{\VAN}[3]{#2}
\let\VANthebibliography\thebibliography
\def\thebibliography{\DeclareRobustCommand{\VAN}[3]{##3}\VANthebibliography}

\usepackage{graphicx}	
\usepackage{amsmath}	
\usepackage{tabularx}
\usepackage[table,xcdraw]{xcolor}
\usepackage{newtxtext,newtxmath}
\usepackage{float}

\newcommand\HI{${\rm H}${\sc i}}
\newcommand{\HII}{\rm H{\sc{ii}}}
\newcommand{\Hmol}{${\rm H}_{2}$}

\newcommand{\eagle}{EAGLE}
\newcommand{\tng}{TNG}
\newcommand{\simba}{SIMBA}

\newcommand{\Mt}[1]{${\mathcal M}^{{\rm T}_{#1}}$}
\newcommand{\logmhalo}{$\log_{10}(M_{\rm 200c}/{\rm M}_{\odot})$\ }

\title[Gas flows in modern cosmological simulations]{The baryon cycle in modern cosmological hydrodynamical simulations}

\author[R. J. Wright et al.]{Ruby J. Wright$^{1,2,3,4}$\thanks{E-mail: ruby.wright@helsinki.fi}, Rachel S. Somerville$^{2}$, Claudia del P. Lagos$^{3,4}$, Matthieu Schaller$^{5,6}$, Romeel Davé$^{7,9}$ , \newauthor{Daniel Anglés-Alcázar$^{8,2}$, Shy Genel$^{2,10}$} \\\ \\ 
$^{1}$Department of Physics, University of Helsinki, Gustaf Hällströmin katu 2, FI-00014 Helsinki, Finland\\
$^{2}$Center for Computational Astrophysics, Flatiron Institute, 162 5$^{\it th}$ Avenue, New York, NY 10010, USA\\
$^{3}$International Centre for Radio Astronomy Research, University of Western Australia, 7 Fairway, Crawley, WA 6009, Australia\\
$^{4}$ARC Centre of Excellence for All Sky Astrophysics in 3 Dimensions (ASTRO 3D)\\
$^{5}$Lorentz Institute for Theoretical Physics, Leiden University, PO Box 9506, NL-2300 RA Leiden, The Netherlands\\
$^{6}$Leiden Observatory, Leiden University, PO Box 9513, NL-2300 RA Leiden, The Netherlands\\
$^{7}$ Institute for Astronomy, University of Edinburgh, Royal Observatory, Edinburgh EH9 3HJ, UK\\
$^{8}$ Department of Physics, University of Connecticut, 196 Auditorium Road, U-3046, Storrs, CT, 06269, USA \\
$^{9}$  Department of Physics and Astronomy, University of the Western Cape, Bellville, Cape Town 7535, South Africa \\ 
$^{10}$ Columbia Astrophysics Laboratory, Columbia University, 550 West 120th Street, New York, NY 10027, USA}

\date{Accepted XXX. Received YYY; in original form ZZZ}

\pubyear{2024}

\begin{document}
\label{firstpage}
\pagerange{\pageref{firstpage}--\pageref{lastpage}}
\maketitle

\begin{abstract}
In recent years, cosmological hydrodynamical simulations have proven their utility as key interpretative tools in the study of galaxy formation and evolution. In this work, we present a comparative analysis of the baryon cycle in three publicly available, leading cosmological simulation suites: \eagle, IllustrisTNG, and \simba. While these simulations broadly agree in terms of their predictions for the stellar mass content and star formation rates of galaxies at $z\approx0$, they achieve this result for markedly different reasons. In \eagle\ and \simba, we demonstrate that at low halo masses ($M_{\rm 200c}\lesssim 10^{11.5}\, M_{\odot}$), stellar feedback (SF)-driven outflows can reach far beyond the scale of the halo, extending up to $2-3\times R_{\rm 200c}$. In contrast, in \tng, SF-driven outflows, while stronger at the scale of the ISM, recycle within the CGM (within $R_{\rm 200c}$). We find that AGN-driven outflows in \simba\ are notably potent, reaching several times $R_{\rm 200c}$ even at halo masses up to $M_{\rm 200c}\approx10^{13.5}\, M_{\odot}$. In both \tng\ and \eagle, AGN feedback can eject gas beyond $R_{\rm 200c}$ at this mass scale, but seldom beyond $2-3\times R_{\rm 200c}$. We find that the scale of feedback-driven outflows can be directly linked with the prevention of cosmological inflow, as well as the total baryon fraction of haloes within $R_{\rm 200c}$. This work lays the foundation to develop targeted observational tests that can discriminate between feedback scenarios, and inform sub-grid feedback models in the next generation of simulations.

\end{abstract}


\section{Introduction}\label{sec:introduction}

The baryon cycle represents a complex and interconnected set of processes, including gas accretion, star formation, chemical enrichment, and large scale outflows, that collectively drive the formation and evolution of galaxies (e.g., \citealt{Oppenheimer2006,Ford2014,Somerville2015,Tumlinson2017,AnglesAlcazar2017b,Oppenheimer2018}). The interplay of these processes establishes a causal link between different scales: the interstellar medium (ISM), circum-galactic medium (CGM), and intergalactic medium (IGM); as well as a connection between the different baryonic components involved at each of these scales: stars, gas, and black holes.  

The cosmological inflow of gas from the IGM, through the CGM, and eventually into the ISM provides the requisite fuel for the process of star formation  (\citealt{Keres2005,SanchezAlmeida2014}). Star formation occurs within the dense regions of the ISM, where efficient cooling and gravitational collapse leads to the initiation of nuclear fusion and the consequent birth of new stars (e.g. \citealt{Kennicutt1983,McKee2007,Kennicutt2012}). A fraction of the gas that makes it into galaxy nuclei can also be accreted onto supermassive black holes (SMBH), which are known to inhabit the dense central regions of many galaxies (see reviews by \citealt{Ferrarese2005,Kormendy2013,Heckman2014}).


Both star formation and accretion onto SMBH drive large scale outflows, which play a crucial role in regulating the baryon cycle (\citealt{Veilleux2005,Veilleux2020}). Stellar feedback, through supernovae explosions, radiation, and stellar winds, injects energy, momentum, and enriched material into the ISM and potentially beyond \citep{Heckman2015}. 
Active galactic nucleus (AGN) feedback, powered by accretion onto supermassive black holes at galactic centers, also releases substantial amounts of energy and momentum, that can drive powerful outflows, impacting the surrounding gas and quenching star formation in some cases (e.g. \citealt{Fabian2012,Heckman2014,King2015}).

On the scale of the ISM, various observational techniques provide relatively well-constrained measurements of the stellar mass, galaxy cold gas (\HI\ and \Hmol) content, and star formation rates. Beyond the ISM, direct observational constraints on the CGM and IGM gas are associated with increasing uncertainty due to its diffuse nature (e.g. \citealt{Werk2014,Tumlinson2017}). Outside the Milky Way, gas accretion has been detected in a number of cases where absorption signatures are identified in galaxy spectra (e.g. \citealt{Rubin2012,Martin2012,Stone2016}; for a review see \citealt{Rubin2017}); and in a number of studies where it is possible to use a fortuitously-aligned background quasar to probe circumgalactic gas flows (e.g. \citealt{Lehner2013,Bouche2016,Martin2019}). Even with these discoveries, the rate at which the gas accretes and its associated properties have yet to be well-quantified on a {\it statistical} basis; and are associated with considerable uncertainties. 


There is a significant body of literature presenting supporting evidence for the presence of feedback-driven gas outflows from  galaxies (e.g., \citealt{Heckman2000,Martin2005,Veilleux2005,Feruglio2010,Newman2012,Rubin2014,Heckman2015,Schroetter2016,Chisholm2016,Rupke2019}; for a recent review see \citealt{Veilleux2020}). However, in each case, the determination of associated mass flux is affected by several systematic uncertainties. In particular, a given outflow tracer only provides information about gas on a subset of spatial scales (nominally confined to well within a given halo virial radius), and to specific gas phases. Furthermore, these measurements are biased towards gas rich galaxies exhibiting particularly strong outflows -- either those at high redshift, or local galaxies experiencing a starburst. 
 
Modern cosmological hydrodynamical simulations constitute a powerful tool to explore the role of different processes within the baryon cycle, and galaxy evolution more broadly (e.g. \citealt{Schaye2010,Dubois2014,Vogelsberger2014,Christensen2016,Schaye2015,Pillepich2018,Hopkins2018,Dave2019}).
However, in order to accurately simulate the formation of galaxy populations within the context of large-scale structure, it is crucial to model essential small-scale processes taking place within individual galaxies. These processes, such as star formation, black hole formation and accretion, and stellar and AGN feedback, operate at scales below the resolution capabilities of the simulations. Therefore, they are commonly incorporated using "sub-grid" techniques, which effectively account for their influence on galaxy evolution in a physically-motivated, but necessarily coarse-grained manner (for reviews, see \citealt{Somerville2015,Vogelsberger2020}). A wide range of sub-grid techniques have demonstrated success in reproducing realistic galaxy populations in cosmological simulations, particularly in capturing the observed shape of the stellar mass function (e.g. \citealt{Thorne2021,Schaye2023}, see also \S\ref{sec:results:1}). 

Despite these promising results, recent studies have suggested that the behavior of the gaseous phase and the broader-scale baryon cycle differ significantly between different simulations and feedback models, and can be degenerate in producing the same stellar mass assembly in galaxies over cosmic time \citep{Naab2017,Mitchell2018,Mitchell2020a,Pandya2020,Dave2020,Villaescusa2021,Ni2023,Tillman2023b}. For example, \citet{Dave2020} compare the cold gas content of galaxies in three publicly available modern cosmological hydrodynamical simulations -- \eagle, TNG100, and \simba\ (for a technical overview of the sub-grid prescriptions in each simulation, see \S\ref{sec:methods}). They find that these three simulations produce relatively similar predictions for the molecular gas content of galaxies (\Hmol\ mass function or \Hmol MF) over cosmic time, within a range of $\lesssim0.3 \,$dex. This phase is strongly correlated with star formation, which also agrees fairly well between models as also established by the aforementioned stellar mass function. Predictions for the slightly more tenuous atomic phase, however, show greater divergence.

Similarly, the emergent gas mass outflow rates and mass loading factors ($\eta\equiv\dot{M}_{\rm out}(r)/\dot{M}_{\star}$) at ISM scales that arise from different sub-grid implementations of stellar-driven winds appear to differ significantly even across simulations that are all calibrated to reproduce the observed $z=0$ stellar mass function. \citet{Mitchell2020a} show a comparison of the ISM-scale mass loadings measured in the \eagle\ simulation and those reported by \citet{Nelson2019} from the TNG50 run of the IllustrisTNG simulation suite. Their results suggest that the gas outflow rates in TNG decline rapidly with increasing radius, such that most of the outflowing ISM material does not make it out of the CGM, while  in \eagle\, outflows are actually powerful enough to leave the CGM and halo entirely -- even entraining some gas on the way. However, we stress that the comparisons between these simulations were not conducted with the same methodology -- \citet{Nelson2019} calculate Eulerian instantaneous flow rates, while \citet{Mitchell2020a} instead use a Lagrangian particle tracking method. In addition, flow quantities have not necessarily been measured at the same physical scales or using the same definitions of halo properties.

To date, a comprehensive examination of the distinct gas cycling paradigms within these simulations has not been carried out. In this paper, we aim to fill this gap by conducting a detailed comparison of the baryon cycle in the \eagle, \tng, and \simba\ simulations \emph{using the same definitions and methodology}. Our primary focus is to gain insight into the physical extent of feedback-driven outflows across a broad range of halo masses. By undertaking such an analysis, we lay the foundation for developing targeted observational tests that can help to assess specific simulation methodologies, and move towards a more constrained understanding of the role of feedback processes in the baryon cycle. 

This paper is arranged as follows: in \S\ref{sec:methods}, we outline each of the simulations -- \eagle, \tng, and \simba\ -- that we analyse in this study, and the uniform methodology we employ in each to measure gas flow rates. In \S\ref{sec:results:1}, we outline the similarities and differences between statically measured baryon content and baryon distribution in haloes in these simulations. In \S\ref{sec:results:2}, we present a detailed analysis of the gas flow rates at different scales surrounding galaxies in these simulations, and how these differences can be used to explain the findings in \S\ref{sec:results:1}. In \S\ref{sec:discussion}, we discuss the implications of our findings forthe baryon cycle, and discuss future targeted observational tests for different feedback models. Finally, we conclude with \S\ref{sec:summary}, where we summarise the main findings of the paper.


\section{Methods}\label{sec:methods}

In this paper, we make use of three modern cosmological hydrodynamical simulations with publicly available data -- namely \eagle, \tng, and \simba. These simulations are all based on a $\Lambda$CDM (cold dark matter) cosmology - in which hierarchically assembled cold dark matter haloes provide the formation site of galaxies; all within the context of an expanding, dark energy dominated universe. As noted above, all of these simulations contain phenomenological sub-grid recipes, which are discussed in more detail below, containing parameters that are calibrated via comparison to observational data or quantities derived from observations. These calibrations are anchored using observations of the stellar mass and/or distribution of galaxies, otherwise leaving the behaviour of the gaseous phase as a ``prediction'' of the models (see \citealt{Crain2023} for an overview). \footnote{An exception to this is the \tng\ AGN feedback model, which is calibrated to reproduce $z\approx0$ halo gas fractions at high mass, and SFR densities over cosmic time. Even so, there is considerable degeneracy in how the exact interplay between gas flows and the baryon cycle can produce these results.}. 



In \S\ref{sec:methods:sims} we introduce the simulations we employ for this study and describe their respective sub-grid feedback implementations. Subsequently, in \S\ref{sec:methods:flows} we outline the method with which we self-consistently measure gas flows rates in each of the simulations.  

\begin{table*}
\centering
\begin{tabular}{c||c|c|c}
\cline{2-4}

& \cellcolor[HTML]{98BDE4}{\color[HTML]{000000} \begin{tabular}[c]{@{}c@{}}\  \\ EAGLE\\  \citet{Schaye2015}\\ \ \end{tabular}}                             & \cellcolor[HTML]{A288CB}{\color[HTML]{000000} \begin{tabular}[c]{@{}c@{}}{ TNG100}\\ \citet{Pillepich2018} \end{tabular}}& \cellcolor[HTML]{E88C8C}{\color[HTML]{000000} \begin{tabular}[c]{@{}c@{}}{ SIMBA}\\ \citet{Dave2019}\end{tabular}}   \\ \hline\hline

\multicolumn{1}{c||}{\begin{tabular}[c]{@{}c@{}} \ \\ Box size \\ \ \end{tabular}} & \begin{tabular}[c]{@{}c@{}} $67.8\,h^{-1}{\rm Mpc}$ (REF)\\ $16.9\,h^{-1}{\rm Mpc}$ (RECAL)\end{tabular}&  $75\,h^{-1}{\rm Mpc}$ &  $50\,h^{-1}{\rm Mpc}$\\ \hline

\multicolumn{1}{c||}{\begin{tabular}[c]{@{}c@{}}\ \\ Element resolution\\ \ \end{tabular}} & \begin{tabular}[c]{@{}c@{}} $1.8 \times 10^6\, {\rm M}_{\odot}$ (REF)\\  $2.3 \times 10^5\, {\rm M}_{\odot}$ (RECAL) \end{tabular} &  $1.4 \times 10^6\, {\rm M}_{\odot}$ &  $1.8 \times 10^7\, {\rm M}_{\odot}$\\ \hline

\multicolumn{1}{c||}{\begin{tabular}[c]{@{}c@{}}\ \\ Hydrodynamics\\ \ \\ \ \end{tabular}}    & \begin{tabular}[c]{@{}c@{}}Smooth particle hydrodynamics (SPH).\\ GADGET-3 + ANARCHY.\\ \citet{Springel2005,Schaller2015}.\end{tabular} & \begin{tabular}[c]{@{}c@{}}Moving Voronoi Mesh (MVM).\\ AREPO (with MHD).\\ \citet{Springel2010}.\end{tabular}                                & \begin{tabular}[c]{@{}c@{}}Meshless Finite Mass (MFM).\\ GIZMO.\\ \citet{Hopkins2015,Hopkins2017}.\end{tabular} \\ \hline

\multicolumn{1}{c||}{\begin{tabular}[c]{@{}c@{}}\ \\ \ \\ Star formation\\ \ \\ \ \end{tabular}} & \begin{tabular}[c]{@{}c@{}}Informed by \citet{Kennicutt1998}.\\ Density \& $Z$ criterion (\citealt{Schaye2004}). \\ \citet{Schaye2008}.\end{tabular}   & \begin{tabular}[c]{@{}c@{}} Informed by \citet{Kennicutt1998}.\\ Density threshold. \\ \citet{Springel2003}.\end{tabular}& \begin{tabular}[c]{@{}c@{}} Informed by \citet{Kennicutt1998}.\\ \Hmol\ -based criterion. \\ \citet{Krumholz2011}.\end{tabular} \\ \hline

\multicolumn{1}{c||}{\begin{tabular}[c]{@{}c@{}}\ \\ \ \\ Stellar feedback\\ \ \\ \ \end{tabular}} & \begin{tabular}[c]{@{}c@{}}\\ Thermal only. \\ Particles heated by $\Delta T_{\rm SF}=10^{7.5}$ K.\\ Energy set by local $\rho$, $Z$. \\ No hydrodynamic decoupling. \\  \citet{DallaVecchia2012}.\\ \ \end{tabular}                              & \begin{tabular}[c]{@{}c@{}}Thermal \& kinetic. \\ 90\% momentum, 10\% thermal.\\ $\eta$ set by local metallicity. \\ $v_{\rm wind}$ set by  $\sigma_{\rm DM}$, $z$.\\ Wind elements decoupled.\\ \citet{Pillepich2018}. \end{tabular} & \begin{tabular}[c]{@{}c@{}}Thermal \& kinetic.\\ 30\% of wind particles injected "hot".\\ $\eta (M_{\star})$ from FIRE \citep{AnglesAlcazar2017b}. \\ $v_{\rm wind}(V_{\rm circ})$ from FIRE \citep{Muratov2015}. \\ Wind elements decoupled. \\ \citet{Dave2019}.\end{tabular} \\ \hline

\multicolumn{1}{c||}{\begin{tabular}[c]{@{}c@{}}\ \\ \ \\ AGN feedback\\ \ \\ \ \\ \ \end{tabular}}& \begin{tabular}[c]{@{}c@{}} \ \\ Accretion: modified Bondi-Hoyle.\\ Feedback: one mode, thermal only. \\ Particles heated by $\Delta T_{\rm AGN}=10^{8.5}$ K.\\ No directionality. \\ No hydrodynamic decoupling. \\ \citet{Schaye2015}.\end{tabular}         & \begin{tabular}[c]{@{}c@{}}\ \\ Accretion: modified Bondi-Hoyle. \\ Low $f_{\rm Edd}$: kinetic feedback.\\ High $f_{\rm Edd}$: thermal feedback. \\ No directionality.\\ No hydrodynamic decoupling.\\ \citet{Weinberger2017}.\end{tabular} & \begin{tabular}[c]{@{}c@{}}\ \\ Accretion: Bondi-Hoyle (hot), torque-limited (cold).\\ Low $f_{\rm Edd}$ (jet-mode): kinetic \& thermal. \\ High $f_{\rm Edd}$ (radiative-mode): kinetic feedback.\\ Outflows injected as bipolar. \\ Jets are decoupled for $10^{-4}\, t_{\rm H}(z)$. \\ \citet{AnglesAlcazar2017a}. \end{tabular}         


\end{tabular}
\caption[]{Description of the simulations used for this study: \eagle, \tng, and \simba. Key parameters include the simulation box size, resolution, hydrodynamics scheme, gas cooling \& star formation prescription, and the implementation of stellar and AGN feedback.}
\label{tab:methods:simtable}
\end{table*}

\subsection{Cosmological simulations}\label{sec:methods:sims}

In Table \ref{tab:methods:simtable} we compare some key features of the simulations used in this paper. We then discuss these features in further detail in \S\ref{sec:methods:sims:eagle}, \ref{sec:methods:sims:tng}, and \ref{sec:methods:sims:simba} for the \eagle, \tng, and \simba\ simulations respectively. Each of these simulations share a base $\Lambda$CDM cosmology, with the choice of cosmological parameters differing by only of order $\approx 1\%$. While the simulations take slightly different approaches to solving the equations describing the hydrodynamics of the gas, they have many characteristics in common: each has a spatial resolution of order $1-10\, {\rm kpc}$, make use of very similar tabulated gas cooling prescriptions, and utilise an effective equation of state (EoS) for cool gas within the ISM. The critical difference between these simulations, for the purposes of this study, is in their approach to modeling the unresolved feedback from star formation (stellar feedback) and SMBHs (AGN feedback), which we focus on below.

\subsubsection{The \eagle\ simulations}\label{sec:methods:sims:eagle}

The \eagle\ simulations, as described by \citet{Schaye2015} and \citet{Crain2015}, utilised a modified version of the N-body Tree-ParticleMesh (TreePM) smoothed particle hydrodynamics (SPH) solver called GADGET-3, initially presented in \citet{Springel2005}. The set of modifications (ANARCHY) are outlined in \citet{Schaller2015}; and include enhancements to the hydrodynamics solver and sub-grid physics modules. The initial conditions for this simulation were generated using the method described in \citealt{Jenkins2013}, assuming the cosmological parameters from \citet{PlanckCollaboration2014}:  $\Omega_{\Lambda} = 0.693$, $\Omega_{\rm m} = 0.307$, $\Omega_{b} = 0.04825$, $H_{0} = 67.77 {\rm km} {\rm s}^{-1} {\rm Mpc}^{-1}$, $\sigma_{8} = 0.8288$, and $n_{\rm s} = 0.9611$.

The flagship box of the suite, referred to as Ref-L100N1504, has a periodic side length of $67.8\, h^{-1}\,{\rm Mpc}$. It was executed with $1,504^3$ dark matter and baryonic particles, resulting in particle masses of $1.8 \times 10^6\, {\rm M}_{\odot}$ and $9.7 \times 10^6 \, {\rm M}_{\odot}$ for baryons and dark matter respectively. The Plummer-equivalent gravitational softening length is set to $2.66 {\rm ckpc}$, limited to a maximum proper length of $0.7 {\rm pkpc}$. Additionally, we present findings from the EAGLE-Recal simulation (Recal-L25N752 in \citealt{Schaye2015}) in Appendices \ref{sec:apdx:densityprofs} and \ref{sec:apdx:tempprofs}. EAGLE-Recal has a box side length of $16.9\,h^{-1}\, {\rm Mpc}$ and achieves twice the spatial resolution with an initial baryonic particle mass of $2.3\times 10^{5}\, {\rm M}_{\odot}$. We present results from the EAGLE-Recal simulation at $z\approx0$ to compare with the behaviour of the fiducial EAGLE model, though the findings we will present in this paper are qualitatively similar for the different resolutions. 

Photoheating and radiative cooling are implemented in \eagle\ based on the work of \citet{Wiersma2009}, including the influence of eleven elements: H, He, C, N, O, Ne, Mg, Si, S, Ca, and Fe \citep{Schaller2015}; with the UV and X-ray background described by \citet{Haardt2001}. Metals are not passively diffused between gas particles in proximity, however cooling calculations are conducted using SPH-kernel weighted metallicities. Since the \eagle\ simulations do not have the resolution to model cold, interstellar gas, a density-dependent temperature floor is imposed (normalised to $T=8\ 000$~K at $n_{\rm H}=10^{-1}{\rm cm}^{-3}$). To model star formation, a metallicity-dependent density threshold is set, above which star formation is locally permitted \citep{Schaye2004}, and gas particles meeting this threshold are converted to star particles stochastically. The star formation rate is set by a tuned pressure law \citep{Schaye2007}, calibrated to the observations of \citet{Kennicutt1983} at $z = 0$. 

The stellar feedback sub-grid model in \eagle\ accounts for energy deposition into the ISM from radiation, stellar winds, and supernova explosions (types Ia and II). This is implemented based on the prescription outlined in \citet{DallaVecchia2012}, via a stochastic thermal energy injection to gas particles in the form of a fixed temperature boost, $\Delta T_{\rm SF} = 10^{7.5}$K. The average energy injection rate from young stars is given by $f_{\rm th}\times8.73\times10^{15}$ erg per gram of stellar mass formed, assuming simple stellar population with a \citet{Chabrier2003} stellar initial mass function (IMF) and that $10^{51}$ erg is liberated per supernova event. The value of $f_{\rm th}$ is a function of local gas density and metallicity, ranging between $0.3-3$ in the fiducial \eagle\ model. The exact parameter choices for the stellar feedback model have been calibrated to observations of the $z\approx0$ galaxy stellar mass function and $M_{\star}-R_{50}$ relation (see \citealt{Crain2015}). 

SMBHs of mass $10^{5}\ h^{-1} {\rm M}_{\odot}$ are seeded when a halo exceeds a virial mass of $10^{10}\ h^{-1} {\rm M}_{\odot}$. Subsequently, SMBHs can grow via Eddington-limited accretion (utilising a Bondi-Hoyle parameterisation), as well as mergers with other SMBHs \citep{Bondi1952,Springel2005a,Schaye2015}. Similar to stellar feedback, AGN feedback in \eagle\ also involves the injection of thermal energy into surrounding particles in the form of a temperature boost of ${\Delta}T_{\rm AGN}=10^{8.5}$K (in the reference physics run; \citealt{Schaye2015}). The rate of energy injection from AGN feedback is determined using the SMBH accretion rate, and a fixed conversion efficiency of accreted rest mass to energy. Importantly, we also note that in \eagle, particles influenced by stellar or AGN feedback are not decoupled from the hydrodynamics when they receive a temperature boost. The choice of temperature boost values are relatively high in order to avoid rapid cooling and dissipation of the feedback energy. The exact parameter choices in the \eagle\ AGN feedback sub-grid model are informed by observations of the $M_{\star}-M_{\rm BH}$ relation \citep{Crain2015}.

\subsubsection{The Illustris\tng\ simulations}\label{sec:methods:sims:tng}
The Next Generation Illustris simulations, known as IllustrisTNG (often shortened to TNG), are a collection of cosmological magneto-hydrodynamical simulations conducted using the moving-mesh refinement code AREPO \citep{Springel2010,Pakmor2011}. Several physics modules in IllustrisTNG were modified relative to its predecessor, Illustris, to achieve better agreement with observations (e.g. \citep{Weinberger2017,Springel2018}. All of the \tng\ boxes assume a \citet{PlanckCollaboration2016} cosmology, with   $\Omega_{\Lambda} = 0.692$, $\Omega_{\rm m} = 0.31$, $\Omega_{b} = 0.0486$, $H_{0} = 67.7 {\rm km} {\rm s}^{-1} {\rm Mpc}^{-1}$, $\sigma_{8} = 0.8159$, and $n_{\rm s} = 0.97$.

The IllustrisTNG suite comprises three primary simulation sets: TNG50, TNG100, and TNG300 \citep{Pillepich2018,Nelson2019}. Each simulation set adopts a different box size and consists of three runs with varying mass resolutions. For our work in this paper, we make use of the TNG100 run with side-length $75.0\,h^{-1}{\rm Mpc}$ to balance mass resolution and volume. The highest resolution TNG100 box, which we adopt for this study, has an initial gas cell mass of $1.40 \times 10^6\, {\rm M}_{\odot}$, and dark matter particle mass of $7.5 \times 10^6 \, {\rm M}_{\odot}$. 
 
Following the original Illustris simulation \citep{Vogelsberger2014}, \tng\ adopts the model of \citet{Springel2003} to model star formation and the pressurization of the ISM. TNG tracks stellar population evolution and chemical enrichment from supernovae type Ia, II, as well as AGB stars, with individual accounting for nine elements (H, He, C, N, O, Ne, Mg, Si, and Fe). Metal-enriched gas undergoes radiative cooling in a redshift-dependent, spatially uniform ionizing UV background radiation field, with corrections for self-shielding in the ISM \citep{Katz1992,Faucher2009}. The cooling contribution from metal lines is included from density, temperature, metallicity, and redshift, following \citet{Wiersma2009} and \citet{Smith2008}. Cooling is further influenced by the radiation field of nearby AGNs, combining the UV background with the AGN radiation field \citep{Vogelsberger2014}. In TNG, diffusion of metals between adjacent gas elements is implemented, modelled based on a gradient extrapolation method \citep{Vogelsberger2014,Pillepich2018}. 

Stellar-driven winds are modeled by isotropically injecting thermal and kinetic energy into wind particles, which are temporarily decoupled from hydrodynamic forces. With 10\% of the feedback energy injected thermally, the kinetic component is parameterised by following prescriptions for mass loading and injection velocity that depend on redshift, gas phase metallicity and the velocity dispersion within a weighted kernel over the 64 nearest dark matter particles  (denoted as $\sigma_{\rm DM}$). A wind particle is recoupled to the gas cell it currently occupies when it either drops below a specific density (5\% of the density threshold for star formation) or reaches a maximum travel time (2.5\% of the current Hubble time). The former criterion is the dominant re-coupling mode, and normally corresponds to a wind particle travelling a distance of a few kpc \citep{Pillepich2018}. 

The black hole growth and feedback physics in \tng\ is described in detail in \citet{Weinberger2017}. When haloes exceed a critical mass of $5\times10^{10}{\rm M}_{\odot}$, a black hole is seeded with an initial mass of $8\times10^5\, h^{-1}{\rm M}_{\odot}$. After seeding, black holes can grow via mergers, or two accretion modes: (i) low accretion state, for which feedback is implemented in the form of kinetic winds; or (ii) the high accretion mode, for which feedback is implemented by depositing thermal energy. The accretion and feedback mode is determined by comparing the accretion rate onto the BH with a critical rate that is black hole mass-dependent. In both of these accretion modes, feedback is injected into surrounding gas cells (the "feedback region") with no preferential direction, and there is no hydrodynamic decoupling of feedback-affected gas elements. 

The calibration process for the TNG model, discussed in \citet{Pillepich2018}, uses various observations such as the stellar mass function, cosmic star formation rate density as a function of redshift, BH mass versus stellar mass relation at $z= 0$, hot gas fraction in galaxy clusters at $z = 0$, and the galaxy mass-size relation at $z = 0$.

\subsubsection{The \simba\ simulations}\label{sec:methods:sims:simba}
\simba\ \citep{Dave2019} is the next generation of the {\sc Mufasa} cosmological galaxy formation simulations \citep{Dave2016}, and use the meshless finite mass (MFM) mode of the GIZMO hydrodynamics code \citep{Hopkins2015,Hopkins2017}, with gravity solver based on GADGET-3 \citep{Springel2005}. The \simba\ simulations adopt cosmological parameters $\Omega_{\Lambda} = 0.70$, $\Omega_{\rm m} = 0.30$, $\Omega_{b} = 0.048$, $H_{0} = 68 {\rm km} {\rm s}^{-1} {\rm Mpc}^{-1}$, $\sigma_{8} = 0.82$, and $n_{\rm s} = 0.97$.

The flagship \simba\ simulation box has a side length of $100\, h^{-1} {\rm Mpc}$, a gas element mass of $1.8\times10^{7}\, {\rm M}_{\odot}$, and DM particle mass of  $9.7\times10^{7}\, {\rm M}_{\odot}$. Due to data storage constraints, we make use of a smaller $50\, h^{-1} {\rm Mpc}$ \simba\ box (with the same element resolution), which we find to provide sufficient statistical power for the current study (m50n1024 in \citealt{Dave2019}).

The simulations track the production of eleven elements (H, He, C, N, O, Ne, Mg, Si, S, Ca, and Fe) originating from Type II and Ia supernovae as well as stellar evolution, with some metals locked away in the process of dust formation. Radiative cooling and photoionisation heating are implemented using the Grackle-3.1 library \citep{Smith2017}. Star formation in \simba\ follows the Kennicutt-Schmidt Law \citep{Kennicutt1998}, scaled by the \Hmol\ fraction calculated for each particle based on local column density and metallicity \citep{Krumholz2011}. Star-formation driven outflows are implemented as decoupled two-phase winds, characterized by updated mass-loading factors derived from particle tracking in the Feedback in Realistic Environments (FIRE) zoom-in simulations \citep{AnglesAlcazar2017b}. Wind velocities are derived from \citet{Muratov2015}. 30\% of wind elements are injected ``hot'', by coupling  thermal energy to the surrounding gas. Wind elements are recoupled when their velocity relative to surrounding gas elements is less than 50\% of the local sound speed. The exact parameter choices in the \simba\ stellar feedback sub-grid model are informed by observations of the $z\approx0$ stellar mass function.

The \simba\ simulations employ two distinct modes for SMBH accretion. One mode involves cold gas supported by rotation, following a gravitational torque model based on \citet{Hopkins2011} and \citet{AnglesAlcazar2017a}. Accretion in this mode is capped at three times the Eddington limit. The second mode applies to hot gas supported by pressure following the standard  Bondi-Hoyle-Lyttleton prescription \citet{Bondi1952}, capped at the Eddington limit. A unique feature of the \citet{AnglesAlcazar2017a} torque-based accretion model is that there is no need to calibrate the AGN feedback efficiency to reproduce the $M_{\star}-M_{\rm BH}$ relation, rather, only an assumption about the accretion efficiency is required. 

\simba\ also incorporates feedback from AGN through radiative and jet modes. The radiative mode drives winds with velocities from $500-1500\, {\rm km}{\rm s}^{-1}$, while the jet mode occurs for SMBHs with Eddington ratios below 0.2 and masses of at least $10^{7.5}\,{\rm M}_{\odot}$. The wind velocity depends on black hole mass (where $v_{\rm wind}=1000\, {\rm km/s}$ for $M_{\rm BH}=10^{9}\, {\rm M}_{\odot}$), and the jet-mode can add a velocity boost of up to $7000\, {\rm km/s}$, meaning outflows can reach speeds of up to $8000\, {\rm km s^{-1}}$. Gas ejected by the jets remains decoupled for a duration that scales with the Hubble time, allowing jets to travel up to $\approx10$ kiloparsecs before the energy is deposited into the surrounding gas. Gas in the jets is injected at the halo virial temperature.

\subsection{Measuring gas flows}\label{sec:methods:flows}
Gas flows across a boundary can be measured in either a Lagrangian or Eulerian manner. In hydrodynamical simulations, the Lagrangian method often involves a gas element tracking scheme whereby the mass of gas elements that are found to have crossed a boundary between two outputs, $t_{i}$ and $t_{\rm f}$, is summed and normalised by $\Delta t=t_{\rm f} - t_{\rm i}$ (e.g. \citealt{Mitchell2020a,Wright2020}). This method is simple to implement in codes where the gas is discretised such that a single gas element represents the same ``parcel'' of gas between snapshots; for instance in the case of \eagle\ (SPH) and \simba\ (MFM). 

The other method for measuring gas flux at a boundary is to take an Eulerian approach, where the instantaneous flow of mass at a boundary can be calculated based on the velocity and mass of boundary elements (e.g. \citealt{Nelson2019}). This method does not require the gas elements to represent the same gas parcel across snapshots as the former method does, it requires only one simulation output as opposed to two, and is more comparable to instantaneous gas flow measurements inferred from observations. As such, we elect to use the Eulerian method for our gas flow calculations, which can be directly applied to the \eagle\ SPH outputs, the \simba/GIZMO MFM outputs, {\it and} the \tng/AREPO outputs (without the need to use tracer particles in the latter case, \citealt{Genel2013}). 

In this work, we focus on calculating gas flow rates across spherical apertures surrounding central galaxies in each of the simulations. We thus make halo catalogues which are derived using a friends-of-friends (FOF) method combined with (i) SUBFIND in the case of \eagle\ and \tng\ \citep{Springel2001,Dolag2009}, or (ii) ROCKSTAR in the case of \simba, to identify central versus satellite subhaloes in all of the simulations. The only characteristics extracted from these catalogues are halo center of potential (which, by definition, is centered on the central galaxy), velocity, and halo mass/radius based on a spherical overdensity of $200\times \rho_c$ ($M_{\rm 200c}$ and $R_{\rm 200c}$ respectively). The rest of the derived baryonic characteristics of haloes are calculated with our own analysis, and we do not expect the small differences in halo finding techniques to affect our results.

At each of the spherical apertures of interest centered at radius $r=R$, we identify boundary gas elements as those residing in a spherical shell of $\Delta r=0.2\times R$, or between $r=0.9\times R$ and $r=1.1\times R$.  We see very little quantitative differences between gas flow rates measured for reasonable choices of $\Delta r$ ($0.05-0.5\times R$, or fixed $\Delta r=1-10 \,{\rm kpc}$). These boundary elements are categorised as being either outflow or inflow depending on the sign of their radial velocity, where the radial velocity of a gas element $i$ relative to a halo center $j$ can be calculated as $v_{{\rm r},\,ij}= \vec{{ v}_{ij}}\cdot \vec{{ r}_{ij}}/\lvert\vec{{ r}_{ij}} \rvert$. The gas flow rates at shell $r=R$ around halo $j$ for each of these subsets of boundary gas elements, $i\in k$, can be calculated with Equation \ref{eq:eulergasflow} as follows:

\begin{equation}
\dot{M}_{k}(r=R)=\frac{1}{\Delta r}\times\sum_{i\in k}\left(m_{i} \frac{{\vec{v}_{ij}}\cdot \vec{{r}_{ij}}}{\lvert \vec{{r}_{ij}} \rvert}\right),
\label{eq:eulergasflow}
\end{equation}

 \noindent{where $\dot{M}_{k}(r=R)$ is the mass flow rate across the spherical boundary at $r=R$. For completeness, in Appendix \ref{sec:apdx:literature} we demonstrate that our method reproduces previous Lagrangian-based gas flow rates in \eagle\ (from \citealt{Mitchell2020a}; Fig. \ref{fig:apdx:literature_mitchell20}), and previous Eulerian-based gas outflow rates in \tng\ (from \citealt{Nelson2019}; Fig. \ref{fig:apdx:literature_nelson19}). }


\begin{figure}
\includegraphics[width=\columnwidth]{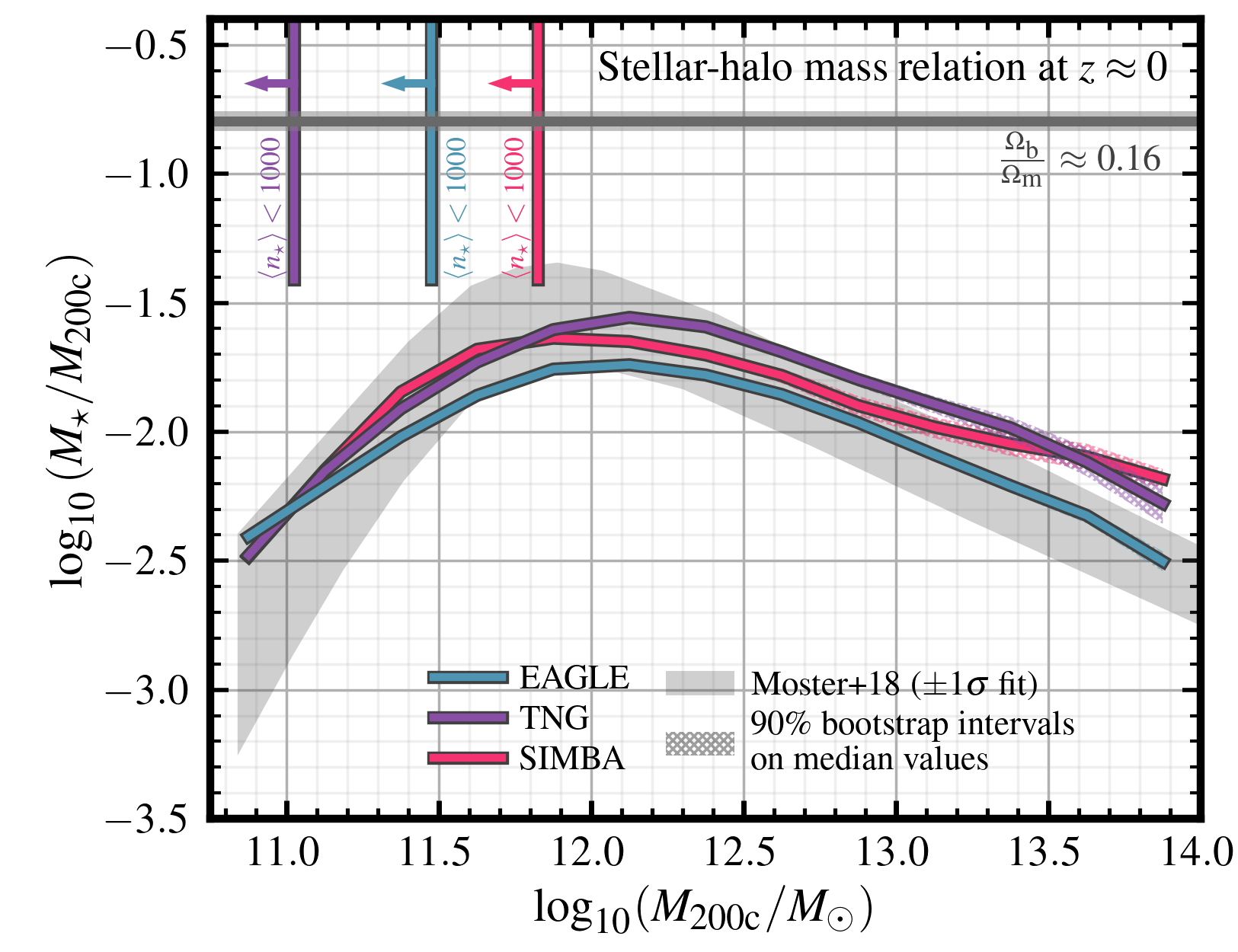}
\caption[]{Median stellar to halo mass fractions as a function of $M_{\rm 200c}$ mass in \eagle\ (blue), \tng\ (purple), and \simba\ (pink) at $z=0$. 
The inferred relation presented in \citet{Moster2018} $\pm 1\sigma$ from their fit is included for reference as a grey shaded region. In each of the simulations, the hatched shaded regions represent the $90\%$ confidence interval on the median from 100 bootstrap re-samplings in each bin. There is general agreement between the simulations and the \citet{Moster2018} relation, despite the simulations each utilising very different sub-grid physics. }
\label{fig:results1:m200_fstar_z0}
\end{figure}

\section{The baryon budget between simulations}\label{sec:results:1}
In this section, before presenting our analysis of gas flows between the simulations in \S\ref{sec:methods:flows},  we compare static properties of the baryons within haloes in \eagle, \tng, and \simba\ at $z=0$. Here, we compare our measurements the baryon content of haloes between simulations, and its breakdown between different components. We note that we reserve a detailed discussion of the link between these findings and previous studies for \S\ref{sec:discussion:haloproperties}. 

Fig.~\ref{fig:results1:m200_fstar_z0} shows the stellar mass relative to the halo mass as a function of halo mass, for central galaxies in each of the three simulations. For comparison, we show the relation derived by \citet{Moster2018}, who use an empirical model that carefully connects observed galaxy properties to simulated dark matter haloes. In \eagle, \tng\ and \simba, the stellar-halo mass ratios are calculated directly from the simulations, including star particles within a $30\, {\rm kpc}$ spherical aperture of the halo centre of potential, and normalising by the halo mass within the $R_{\rm 200c}$ radius, $M_{\rm 200c}$. We convert the virial masses used as the x-values in \citet{Moster2018} from the \citet{Bryan1998} overdensity criterion to the $M_{\rm 200c}$ definition we use here with {\sc hmfcalc} \citep{Murray2013}, however note that we do not adjust the \citet{Moster2018} $f_{\star}$ values at a given halo mass. We also note that the observations that constrain the \citet{Moster2018} empirical model are diverse, and subject to systematic uncertainties and projection effects which we do not attempt to mimic with our simulation-based measurements. Despite these differences in methodology,  Fig.~\ref{fig:results1:m200_fstar_z0} illustrates that each simulation reproduces the reduced galaxy formation efficiency at low- and high- masses required to match observationally derived stellar mass functions (see also \citealt{Somerville2015,Dave2020}). As noted above, and discussed in \citep{Somerville2015}, the relatively good agreement of the different simulations in spite of the very different implementation of sub-grid physics is largely due to the fact that a high weight is placed on reproducing these observational constraints when calibrating the sub-grid parameters.

\begin{figure*}
\includegraphics[width=1\textwidth]{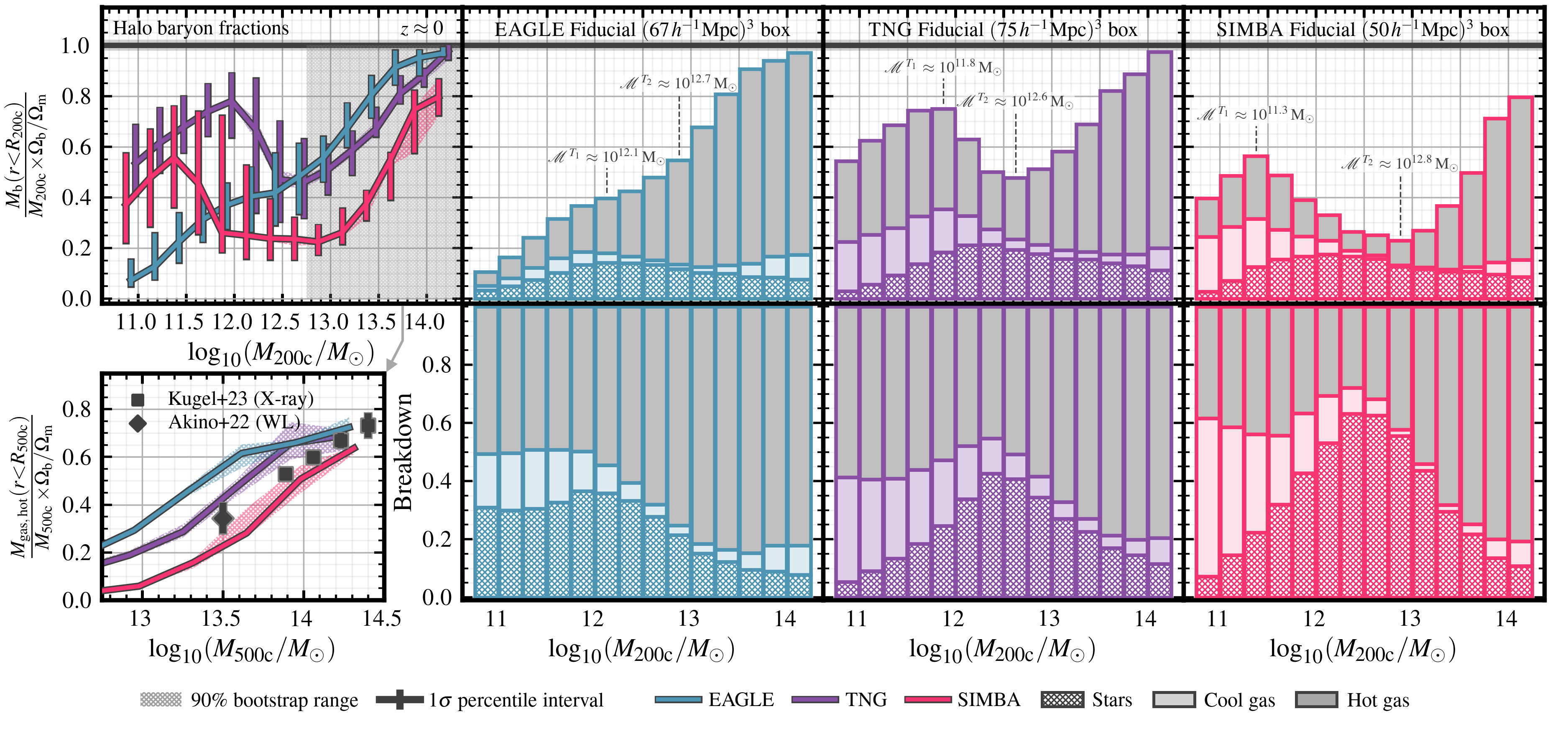}
\caption[]{The baryon content of haloes at $z\approx0$ in \eagle\ (blue), \tng\ (purple) and \simba\ (pink). {\it Top panel, $1^{\rm st}$ column}: The median total baryon fraction within $R_{\rm 200c}$ (stars, gas, black holes) of haloes as a function of $M_{\rm 200c}$ mass. {\it Bottom panel, $1^{\rm st}$ column}: The gas fraction of haloes within $R_{\rm 500c}$ as a function of $M_{\rm 500c}$, shown in comparison to weak lensing observations from \citet{Akino2022} (diamond point); and X-ray data compiled in \citet{Kugel2023} (square points). {\it Top panels,  $2^{\rm nd}-4^{\rm th}$ columns}: Total baryon content for each simulation within $R_{\rm 200c}$ broken down into stars (hatched bars), ``cold'' gas (coloured shaded bars), and ``hot gas'' (grey bars) in each simulation. In the top panels, this baryon content is normalised by $f_{\rm b}\times M_{\rm 200c}$. {\it Bottom panels, $2^{\rm nd}-4^{\rm th}$ columns}: These panels shows the same breakdown, but re-normalised by the actual baryonic mass in each halo, $M_{\rm bar} (r<R_{\rm 200c})$.  In each of the simulations in the left-hand panel, error-bars represent the $16^{\rm th}-84^{\rm th}$ percentiles in halo baryon fractions for a given mass bin; the hatched regions represent the $90\%$ confidence interval on the median from 100 bootstrap re-samplings in each bin. While the simulations produce similar results for the stellar content of haloes at $z\approx0$, the same cannot be said for the gaseous phase. }
\label{fig:results1:m200_scalings_z0}
\end{figure*}

In Fig. \ref{fig:results1:m200_scalings_z0}, we compare the total baryon content of haloes at $z=0$ as a function of halo mass (top left hand panel), and in the remaining columns show how this baryon content is distributed between stars, cold gas, and hot gas in each of the simulations (as a fraction of total halo mass in the top panels, and as a breakdown of the total baryon content in the bottom panels). In each simulation, we define `` cold gas'' as the gas that is either considered star-forming, or below $5\times10^{4} {\rm K}$, and ``hot gas'' as the remaining gas within $R_{\rm 200c}$ that does not meet this criteria. We do not illustrate the contribution from black holes in the breakdown panels as this component is typically very small. In the case of haloes with several galaxies within $R_{\rm 200c}$, we include the contribution from satellite galaxies in calculating the total baryonic mass in each reservoir. Also, unlike Fig. \ref{fig:results1:m200_fstar_z0}, in this plot the stellar content of the halo is determined using all stellar particles within $R_{\rm 200c}$, as opposed to the central $30{\rm kpc}$. This means that in Fig. \ref{fig:results1:m200_scalings_z0}, the stellar component also includes contributions from intra-halo stars and satellite galaxies within $R_{\rm 200c}$. The ``cold gas'' component is not decomposed further or post-processed to provide \HI\ and \Hmol\ content -- for a direct comparison of galaxy neutral hydrogen abundances in each of the simulations, we refer the reader to \citet{Dave2020}. Furthermore, a phase decomposition of CGM gas as a function of halo mass is also presented for \simba\ in \citet{Sorini2022}.

One of the striking differences between the simulations in terms of halo baryon fraction is their behaviour in the range \logmhalo$\in[11.5,  12.5]$. \simba\ and \tng\ roughly exhibit the same functional form in this mass range -- halo baryon fractions increase with halo mass up to a turnover mass which we denote as \Mt{1}. Above \Mt{1}, halo baryon fractions decrease with halo mass until a second turnover mass, denoted by \Mt{2}, is reached. Above \Mt{2}, halo baryon fractions steadily rise again to the highest halo masses presented here at \logmhalo$=$14. Physically, these transitions produce three distinct ``regimes'' in terms of the $M_{\rm 200c}-f_{\rm b}$ relation: (i) sub-\Mt{1}, where baryon fractions increase with halo mass as a result of the decreasing efficiency of stellar feedback in removing gas from a deeper potential well; (ii) the transition range \Mt{1} to \Mt{2}, where this trend reverses as AGN feedback grows in importance, and (iii) high mass $>$\Mt{2}\ haloes where similar to regime (i), the efficiency of AGN feedback in removing baryonic material decreases in very massive group-cluster mass haloes. We note that in the case of \eagle, in the \Mt{1}-\Mt{2}\ range, there is only a slight flattening in halo baryon fractions with increasing mass, as opposed to a clear turnover. These halo baryon fractions, as well as the behaviour of simulation runs without AGN feedback (shown in Fig. 8 in \citealt{Pillepich2018} for TNG, and Fig. 7 in \citealt{Correa2018a} for EAGLE) indicate that AGN activity is responsible for this turnover/flattening behaviour respectively between \Mt{1}-\Mt{2}. 

At halo masses below \logmhalo$\approx11.5$, corresponding to sub-\Mt{1}\ masses, the left hand panel of Fig. \ref{fig:results1:m200_scalings_z0} demonstrates that all of the simulations presented have total baryon fractions that increase with halo mass. While this behaviour with halo mass is the same, the baryon fractions themselves are markedly different. In this mass regime, the fiducial L100-REF \eagle\ model predicts the lowest baryon fractions of all the simulations presented. At a halo mass of \logmhalo$\approx$11, the L100-REF model predicts a median halo baryon fraction of $\approx15\%$ of the universal baryon fraction ($\Omega_{\rm b}/\Omega_{\rm m}\approx0.16$). While we do not display the results here, we note that in the L25-RECAL model at this mass, halo baryon fractions are slightly higher at $\approx30\%$, though still lower than the corresponding baryon fraction predictions from \simba\ at $\approx40\%$ of the expected cosmological baryon fraction. \tng\ predicts the highest halo baryon fractions at this mass, at $\approx60\%$ of the universal value (see also \citealt{Davies2019a,Davies2019b}). 

In this sub-\Mt{1}\ mass regime, we also note significant differences in the way that this baryonic matter is partitioned between the simulations. In the fiducial \eagle\ model, the low overall baryon fractions mean that in order to match the expected stellar content of haloes, stars make up over $25\%$ of the baryonic content, compared to \simba\ and \tng\ where the stellar component only represents $\lesssim10\%$ of the baryonic matter. For halo masses below \logmhalo$\approx11.5$, we note that the \eagle\ L100-REF run predicts very low halo cold gas content -- always less than half of the predicted stellar content. This is not the case in \tng\ and \simba, where the stellar to cold gas mass ratio is near unity in this mass range. \citet{Dave2020} show that the \eagle\ fiducial run underpredicts the observed $z=0$ \HI MF from ALFALFA observations by $\approx0.5\, {\rm dex}$ \citep{Jones2018}; but matches the local CO luminosity function from \citet{Saintonge2017} fairly well (see also \citealt{Lagos2015}). This suggests that the fiducial \eagle\ run lacks the gas associated with the more diffuse atomic phase at $z=0$, however this result is heavily dependent on how gas phases are partitioned in post-processing. While not directly shown in this Figure, we remark that the L25-RECAL \eagle\ model produces a better match to the local \HI MF; and that for $z\geq1$, that the \eagle\ L100-REF and L25-RECAL models are consistent with regard to predicted \HI\ and \Hmol\ mass functions. 

While \simba\ and \tng\ predict a similar functional form for baryon fractions across halo  mass, their normalisation and the location of transition masses \Mt{1}\ and  \Mt{2}\ are markedly different (as also found in \citealt{Ni2023,Delgado2023,Crain2023}). In \tng, the first transition mass occurs at $\log_{10}($\Mt{1}$/{\rm M}_{\odot})\approx11.8$; while in \simba, the transition occurs at much lower halo mass, $\log_{10}($\Mt{1}$/{\rm M}_{\odot})\approx11.3$. Furthermore, in \tng\ the second transition mass occurs at $\log_{10}($\Mt{2}$/{\rm M}_{\odot})=12.6$; while in \simba, halo baryon fractions do not increase until $\log_{10}($\Mt{2}$/{\rm M}_{\odot})\approx12.8$. In this transition range, halo baryon fractions in \tng\ are, on average, 20\%-50\% higher than those measured in \simba. Furthermore, at \logmhalo$=$14, baryon fractions in \simba\ only reach $\approx$70\% of $\Omega_{\rm b}/\Omega_{\rm m}$, while haloes in \tng\ haloes approach this universal value very closely. Instead of a turnover, \eagle\ predicts a flattening in halo baryon fractions between $\log_{10}($\Mt{1}$/{\rm M}_{\odot})\approx 12.1$ and $\log_{10}($\Mt{2}$/{\rm M}_{\odot})\approx12.7$. In this transition range, the drop in \tng\ baryon fractions meets the slowly rising \eagle\ baryon fractions at the \Mt{2}\ transition point. Above \Mt{2}, \eagle\ and \tng\ agree very well in terms of predicted baryon fractions as a function of halo mass. 

In the high-mass regime, it is possible to compare predictions for halo gas content with observations. In \citet{Kugel2023}, a number of carefully compiled X-ray and weak lensing (WL) observations were used to calibrate a large-volume cosmological simulation, FLAMINGO.  Such observational datasets are nominally cast as (hot) gas fractions within $R_{\rm 500c}$, as a function of $M_{\rm 500c}$, representing a slightly shrunken aperture relative to the rest of our measurements. In the bottom left panel of Figure \ref{fig:results1:m200_scalings_z0}, we present like-for-like measurements of hot halo gas content in each of the simulations within $R_{\rm 500c}$, and compare with this set of observations compiled in \citet{Kugel2023}. We refer the reader to this study for a detailed explanation of the curated dataset, noting that we only make use of the WL data from \citet{Akino2022}, and that the relevant X-ray datasets come from the following studies: \citet{Vikhlinin2006,Maughan2008,Rasmussen2009,Sun2009,Pratt2010,Lin2012,Lagana2013,Sanderson2013,Gonzalez2013,Lovisari2015,Pearson2017,Lovisari2020}. As above, we define ``hot'' gas as any gas elements with a temperature $T\geq 5\times10^{4}$. We note that this choice of threshold does not significantly influence gas fractions in this mass range, as the majority of gas is at or above halo virial temperature (see top panels). Regardless, we stress that this selection is unlikely to exactly mimic the gas accounted for in the observations, however allows us to understand the difference between simulation predictions and their broad agreement with observations. Our measurements between $M_{\rm 500c}=10^{13.5}{\rm M}_{\odot}-M_{\rm 500c}=10^{14.5}{\rm M}_{\odot}$ indicate that median halo gas fractions in \tng\ and \eagle\ are slightly higher than the compiled observations (more so in \eagle), and slightly lower than observations in \simba. The range of predictions from the simulations is largest between $M_{\rm 500c}=10^{13}{\rm M}_{\odot}-M_{\rm 500c}=10^{14}{\rm M}_{\odot}$, while at higher masses ($M_{\rm 500c}\approx10^{14.5}{\rm M}_{\odot}$), the gas fractions tend towards a tighter range. For reference, we note that the offset between $M_{\rm 200c}$ and $M_{\rm 500c}$ halo mass measurements is relatively constant at $\approx0.15$ dex in this mass range. While the X-ray datasets extend to higher masses, there is a scarcity of cluster-mass haloes in the simulations to used for this study due to their box size.

In Appendix \ref{sec:apdx:fbz2}, we show a breakdown of the baryon content of haloes at $z=2$, finding that the qualitative behaviour of the simulations in relation to baryon fraction across halo mass remains similar to the results presented here at $z=0$, with a universal shift to higher halo baryon fractions. The transition masses \Mt{1}\ and \Mt{2}\ remain similar in \eagle\ and \tng\ at $z=2$ compared to $z\approx0$ galaxies; though this is not the case for \simba. In \simba, the transition masses are higher than at $z=0$, with $\log_{10}($\Mt{1}$/{\rm M}_{\odot})=12.2$ and $\log_{10}($\Mt{2}$/{\rm M}_{\odot})=12.8$ at $z=2$ (see also Fig. 7 in \citealt{Sorini2022}). 

To aid in the interpretations of the gas flow measurements in the following section (\S\ref{sec:results:2}), in Appendices \ref{sec:apdx:densityprofs} and \ref{sec:apdx:tempprofs} we provide radial density and temperature profiles of central galaxies in each of the simulations. Together with the figures contained in this section, these radial profiles provide additional information about the spatial distribution and properties of gas within haloes in each of the simulations.

\section{Gas flows through the ISM, CGM, and IGM}\label{sec:results:2}
In this section, we aim to identify the physical mechanisms behind the differences in the baryon budget predicted by the \eagle, \tng\ and \simba\ simulations as explored in \S\ref{sec:results:1}. Our method for measuring gas flow rates is detailed in \S\ref{sec:methods:flows}. We note that for the purposes of this analysis, we do not enforce any additional selection cuts to classify elements as inflow or outflow other than the sign of their radial velocity.

\subsection{Ejective feedback: gas \textit{\textbf{outflow}} rates as a function of scale}\label{sec:results:2:outflows}

\begin{figure}
\includegraphics[width=\columnwidth]{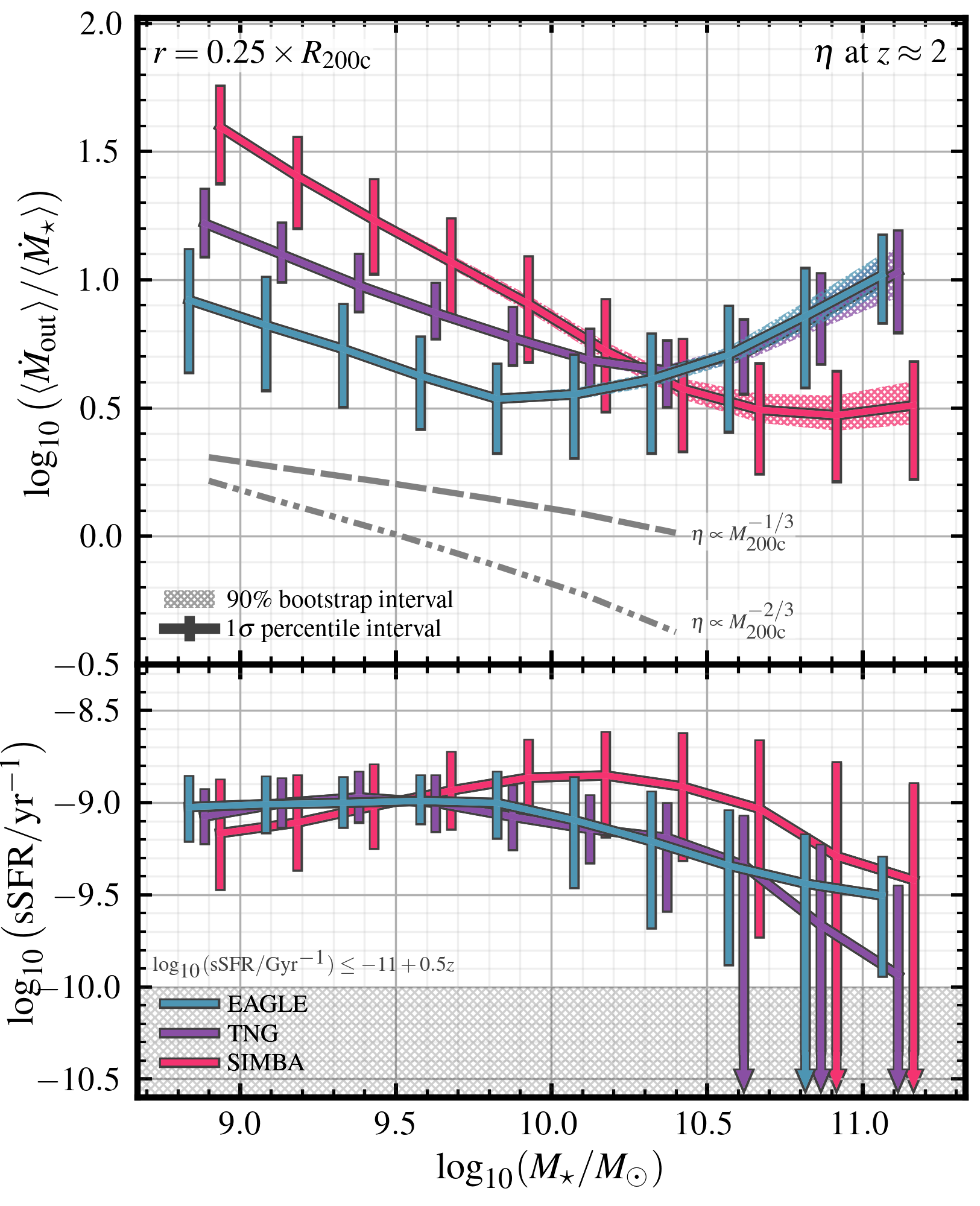}
\caption[]{Mass loading ($\eta$) and specific star formation rates (${\rm sSFR}$) at $z\approx2$ as a function of stellar mass in \eagle\ (blue), \tng\ (purple) and \simba\ (pink). {\it Top panel}: averaged ISM-scale ($r=0.25R_{\rm 200c}$) mass loading (see text for definition) as a function of stellar mass. {\it Bottom panel}: mean specific star formation rate as a function of stellar mass in each of the simulations. In the top panel, the error-bars displayed correspond to the range of measured mass loading values based on the $16^{\rm th}$ and $84^{\rm th}$ percentiles of mass outflow rate in a given stellar mass bin. In the bottom panel, the error-bars correspond to the $16^{\rm th} - 84^{\rm th}$ percentile range of ${\rm sSFR}$ at a given mass. In both panels, coloured hatched regions correspond to the bootstrap-generated $90\%$ confidence interval on the respective averages at a given mass. $\eta$   scales negatively with $M_{\star}$ for all simulations below $M_{\star}\approx10^{10}{\rm M}_{\odot}$, however the $\eta$ values in this mass range can vary between simulations by $\gtrsim 0.5\, $dex on average. }

\label{fig:results2:mstar_etaism_z2}
\end{figure}

In Fig. \ref{fig:results2:mstar_etaism_z2}, we display measurements of galaxy outflow rates at the ISM scale in a form commonly presented in previous literature: the ISM scale mass loading factor, $\eta=\dot{M}_{\rm out}/{\rm SFR}$, where $\langle\dot{M}_{\rm out}\rangle$ is the mean gas outflow rate, and $\langle\dot{M}_{\star}\rangle$ is the mean star formation rate.  We begin by comparing the mass loading results between simulations at $z\approx2$, which builds a picture for the operation of gas flows at the epoch of maximum star formation rate density \citep{Madau2014}. In the bottom panel of Fig. \ref{fig:results2:mstar_etaism_z2}, we also compare the median specific star formation rate ($\dot{M}_{\star}/M_{\star}$) of galaxies. This assists in understanding the influence of star formation rate on the mass loading measurements provided in the top panel. In the bottom panel, we also include a hatched grey region which represents an indicative ``quenched'' region where the specific star formation rate drops below $\log_{10}({\rm sSFR}/{\rm yr^{-1}})={-11+0.5z}$ \citep{Furlong2015}. 

We define the stellar mass of a galaxy in the same way as Fig. \ref{fig:results1:m200_fstar_z0}, including all star particles within $30{\rm \, kpc}$ of the halo centre of potential. Similarly, the star formation rates we use in calculating mass loading are evaluated with gas within $30{\rm kpc}$ of the galaxy centre, which is fixed regardless of the scale that we measure outflow rates. We define the ``ISM'' scale by measuring gas flows across a spherical boundary at $r=0.25\times R_{\rm 200c}$, which we consider to be the outer scale of the ISM. We stress that we do not enforce a minimum radial velocity for outflow gas; meaning that while we do select slow-moving outflow gas elements/particles, these outflow values are fair to compare between the simulations and represent the true mass flux at each scale\footnote{Through experimentation, we find that enforcing a velocity cut slightly reduces the normalisation of mass loadings, but does not change the functional form with galaxy mass, nor the trends between the simulations.}. Furthermore, for the mass flow rate measurements at the scale of the ISM, we do not require the gas to be ``cold'' to be included in the calculation. This choice is made to simplify subsequent comparisons with gas flows at larger scales, where in many cases the majority of the gas has been heated to near the virial temperature of the halo (see Appendix \ref{sec:apdx:tempprofs}).

We follow the approach of \citet{Mitchell2020a} (based on arguments presented in \citealt{Neistein2012}) in calculating mass loading: instead of comparing $\eta\equiv\dot{M}_{\rm out}/{\rm SFR}$ for each halo individually, we calculate $\eta$ for each halo mass as the {\it mean} outflow rate in each halo mass bin divided by the {\it mean} star formation rate in the bin. \citet{Neistein2012} demonstrate that this method correctly predicts the average rate of mass exchange when integrated over time, and also mitigates the impact of discreteness noise that would otherwise influence the results. We note that the percentile ranges displayed in the top panel of Fig. \ref{fig:results2:mstar_etaism_z2} correspond to the range of mass loading values predicted based on the $16^{\rm th}$ and $84^{\rm th}$ percentiles of mass outflow rate in a given stellar mass bin, without including variance in star formation rates at a given mass.

For reference, in Fig. \ref{fig:results2:mstar_etaism_z2}, we also include indicative $\eta$ scalings with $V_{\rm c}\equiv \sqrt{G M_{\rm 200c}/R_{\rm 200c}}$ (halo circular velocity): a ``momentum conserving'' scenario with $\eta\propto V_{\rm c}^{-1}\propto M_{\rm 200c}^{-1/3}$; and an ``energy conserving'' scenario with $\eta\propto V_{\rm c}^{-2}\propto M_{\rm 200c}^{-2/3}$ as dashed and dash-dotted lines respectively (see \citealt{Murray2005} for derivation).

\begin{figure*}
\includegraphics[width=\textwidth]{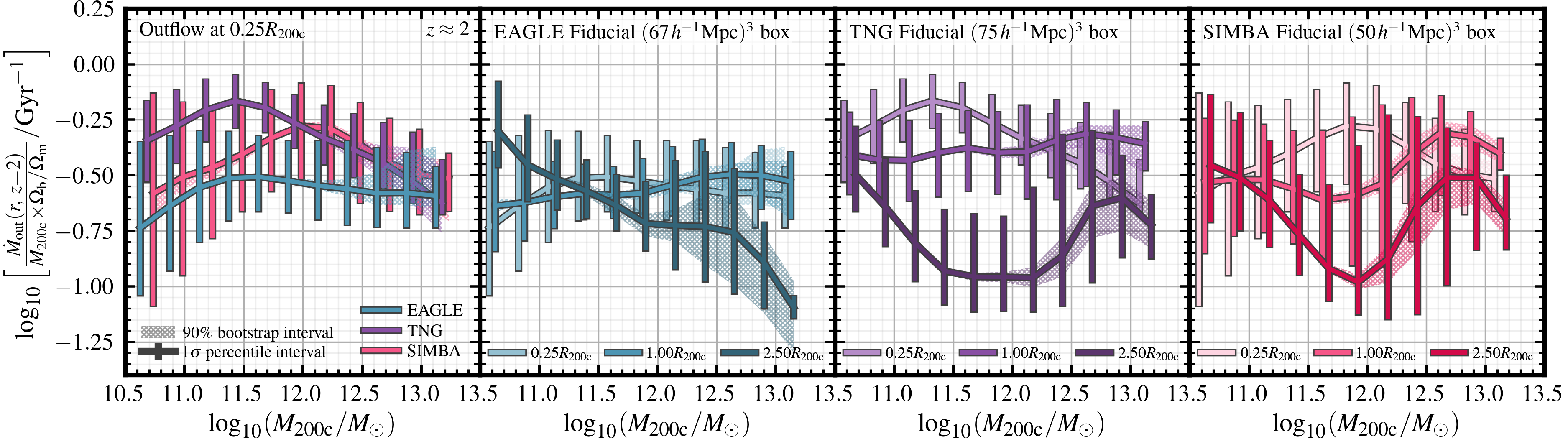}
\caption[]{Mass outflow rates as a function of $M_{\rm 200c}$ at $z\approx2$ in \eagle\ (blue), \tng\ (purple) and \simba\ (pink). Mass outflow rates are normalised by $f_{\rm b}\times M_{\rm 200c}$ in each case to indicate gas outflow ``efficiency'' across  different mass scales. {\it Left panel}: Outflow efficiency at the ISM scale, compared between the simulations. {\it Panels 2-4}: Outflow efficiencies at different scales in \eagle, \tng, and \simba\ respectively. In each of these panels, ISM-scale outflows are plotted with lighter colours, halo-scale outflowsare plotted in medium colours, and larger-scale outflows are plotted in darker colours. In all panels, error-bars correspond to the $16^{\rm th} - 84^{\rm th}$ percentile range in outflow rates at a given mass, and hatched regions correspond to the bootstrap-generated $90\%$ confidence interval on the medians at a given mass. SF-driven outflows reach the greatest scale in \eagle, while AGN-driven outflows reach the greatest scale in \simba. }
\label{fig:results2:m200_outflow_scales_z2}
\end{figure*}

\begin{figure*}
\includegraphics[width=\textwidth]{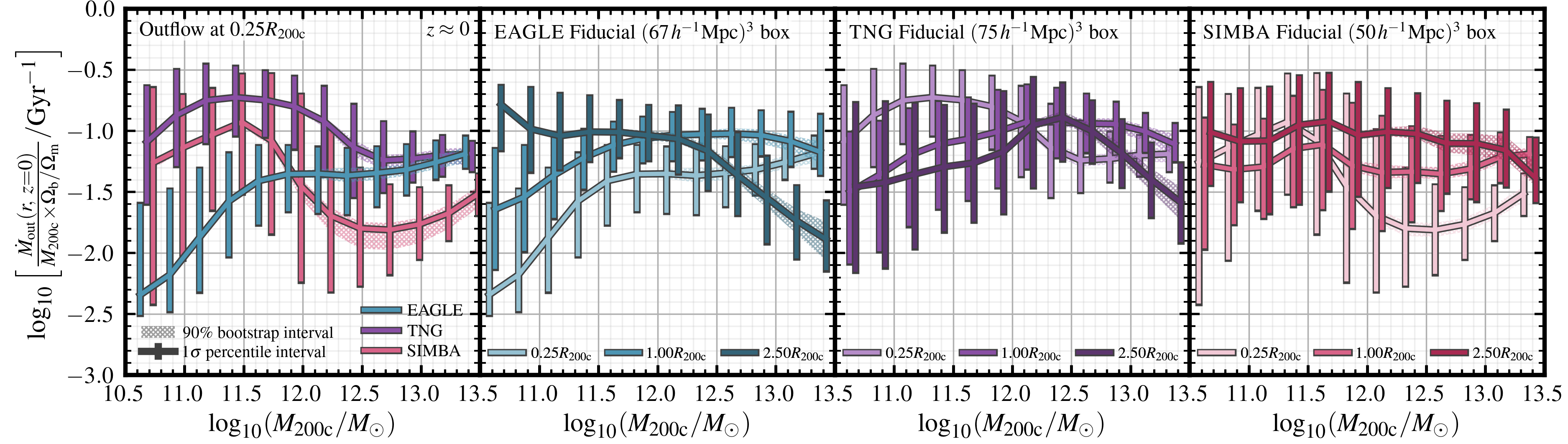}
\caption[]{Mass outflow rates as a function of $M_{\rm 200c}$ at $z\approx0$ in \eagle\ (blue), \tng\ (purple) and \simba\ (pink). Mass outflow rates are normalised by $f_{\rm b}\times M_{\rm 200c}$ in each case to indicate gas outflow ``efficiency'' across  different mass scales. {\it Left panel}: Outflow efficiency at the ISM scale, compared between the simulations. {\it Panels 2-4}: Outflow efficiencies at different scales in \eagle, \tng, and \simba\ respectively. In each of these panels, ISM-scale outflows are plotted with lighter colours, halo-scale outflowsare plotted in medium colours, and larger-scale outflows are plotted in darker colours. In all panels, error-bars correspond to the $16^{\rm th} - 84^{\rm th}$ percentile range in outflow rates at a given mass, and hatched regions correspond to the bootstrap-generated $90\%$ confidence interval on the medians at a given mass. \eagle\ predicts efficient SF-driven ejection of gas from haloes, while \tng\ and \simba\ predict an intra-halo ``recycling'' scenario for SF-affected gas. AGN-driven outflows reach the greatest scale in \simba. }
\label{fig:results2:m200_outflow_scales_z0}
\end{figure*}

 At the ISM-scale of $r=0.25R_{\rm 200c}$ in the top  panel of Fig. \ref{fig:results2:mstar_etaism_z2}, it is clear that each of the simulations predicts a similar negative scaling in $\eta=\dot{M}_{\rm out}/{\rm SFR}$ as a function of $M_{\star}$, until a turnover mass where outflows begin to pick up relative to star formation rates (also demonstrated in \citealt{Nelson2019} and \citealt{Mitchell2020a} for TNG and \eagle\ respectively, though using different scales and methodologies). The turnover mass differs slightly between simulations -- in \eagle\ and \tng, the turnover occurs at $M_{\star}\approx10^{10}{\rm M}_{\odot}$ and $M_{\star}\approx10^{10.3}{\rm M}_{\odot}$ respectively; with $\eta$ values at the turnover masses reaching a minimum of $\eta\approx 10^{0.5}$ and $\eta\approx 10^{0.7}$ respectively. In \simba, the turnover is more gradual, with the minimum average mass loading occurring closer to $M_{\star}\approx10^{11}{\rm M}_{\odot}$. Folding in the stellar-halo mass relation in each of the simulations, in \eagle\ and \tng\ the turnover mass occurs at \logmhalo$\approx11.8-12$, while in \simba, the turnover occurs at a slightly higher mass of \logmhalo$\approx12.4$ at $z\approx2$ (as is also the case for halo baryon content). 

Below the turnover mass, the scaling of $\eta$ with mass for each of the simulations falls between the expected ``momentum conserving'' ($\eta\propto M_{\rm 200c}^{-1/3}$) and ``energy conserving'' ($\eta\propto M_{\rm 200c}^{-2/3}$) scaling relations. The relation is steepest in \simba, for which $\eta$ approximately follows $M_{\rm 200c}^{-2/3}$, and is closer to $M_{\rm 200c}^{-1/2}$ in \eagle\ and \tng. The mass outflow rate from the ISM per unit star formation rate in this mass range is highest in \simba, which predicts $\eta=10^{1.6}$ at $M_{\star}\approx 10^{9} {\rm M}_{\odot}$, followed by \tng\ predicting approximately one third of this outflow rate at $\eta=10^{1.2}$, and \eagle\ predicting approximately half of the \tng\ outflow rates at $\eta=10^{0.9}$, for the same stellar mass. 


The higher mass loading factors predicted in \tng\ and \simba\ at this scale are likely a result of the combination of input mass loading prescriptions, velocities, and the temporarily decoupled nature of these wind gas elements. Even though the wind gas elements in both simulations are likely to have recoupled before reaching the scale where we measure outflow rates ($0.25\times R_{\rm 200c}$), the ability of these gas elements to escape the densest regions of the ISM without losing as much momentum means that the mass flux at this scale can remain quite high. \citet{Nelson2019} show in relation to the \tng50 simulation that while the velocity and $\eta$ of SF-driven feedback are prescribed at injection, the way in which these outflows propagate to larger scales is non-trivial. The mass loading factors presented at the ISM scale agree with the measurements at $10\, {\rm kpc}$ presented in \citet{Nelson2019}, and the \simba\ mass loading at this scale agrees with the input mass loading prescription from \citet{AnglesAlcazar2017b} taken from the FIRE simulations. In comparison, in \eagle, thermal SF-driven outflows do not have prescribed velocities or mass loading. The feedback energy is deposited purely thermally, and the large temperature boosts ($\Delta T_{\rm SF}=10^{7.5}\, {\rm K}$) and associated pressure produce strong outflows. 

Above the mass turnover where AGN feedback becomes the dominant ejection mechanism, there are further clear differences between mass loading in the simulations. \tng\ demonstrates the steepest upturn in ISM-scale mass loading with halo mass, reaching $\eta=10^{1.7}$ at \logmhalo$=13$. At this same mass, \eagle\ predicts a mass loading of approximately half the \tng\ value at $10^{1.4}$, and \simba\ predicts  mass loading at one tenth of the \eagle\ value at $10^{0.6}$ (corresponding to the increased transition mass at which AGN feedback becomes efficient as per Fig. \ref{fig:results1:m200_scalings_z0}). The lower $\eta$ values at high mass in SIMBA are likely linked to the onset of jet-mode feedback, which injects more energy than the QSO-mode feedback with the same momentum flux, corresponding to a much lower mass loading factor relative to the BH accretion rate. 

Comparing the median specific star formation rates in the bottom panel of Fig. \ref{fig:results2:mstar_etaism_z2}, we note that there is broad agreement between the simulations in terms of the shape of the star formation main sequence. There are, however, small quantitative differences between the simulations -- for instance, the median specific star formation rate in \simba\ at $M_{\star}\approx 10^{9} {\rm M}_{\odot}$ sits $0.1-0.2$ dex below \eagle\ and \tng, and sits $0.3-0.4$ dex above \eagle\ and \tng\ for $M_{\star}\approx 10^{10.2}-10^{10.7} {\rm M}_{\odot}$. Together, these slight differences in star formation activity complicate comparisons between outflow behaviour in the simulations using $\eta$. Furthermore, instantaneous measurements of mass loading do not account for the delay between star formation and the associated feedback. As such, for the remainder of this paper, we quote flow rates {\it without} normalizing by star formation rate.  Additionally, citing the (slight) differences in the stellar-halo mass relation between simulations as presented in Fig. \ref{fig:results1:m200_fstar_z0}, we choose to compare the baryon cycle between simulations as a function of halo mass rather than stellar mass for the remainder of this work. Given the very similar underlying cosmological parameters, the form of the halo mass functions in the simulations are very closely matched; meaning that using halo mass as the dependent variable allows us to fairly compare the simulations at a set depth of gravitational potential well. 

In Fig. \ref{fig:results2:m200_outflow_scales_z2} and \ref{fig:results2:m200_outflow_scales_z0}, we show the raw outflow rates for central galaxies in each simulation as a function of halo mass, normalised by $\Omega_{\rm b}/\Omega_{\rm m}\times M_{\rm 200c}$ at $z=2$ and $z=0$ respectively (``outflow efficiency''). In the left hand panels, we show these outflow rates at the ISM scale ($0.25\times R_{\rm 200c}$). In the remaining panels for each simulation, we show outflow rates at the ISM scale, the halo scale (defined at $1.00\times R_{\rm 200c}$), and at the IGM scale (defined at $2.50\times R_{\rm 200c}$).

The values of outflow efficiency indicate the inverse of the time-scale that would be required for a gas mass of $M_{\rm 200c}\times\ \Omega_{\rm b}/\Omega_{\rm m}$ to be removed at this scale. In the left-hand panels of Fig.  \ref{fig:results2:m200_outflow_scales_z2} and \ref{fig:results2:m200_outflow_scales_z0}, we compare median outflow rates in each of the simulations at the ISM scale ($0.25\times R_{\rm 200c}$). In each of the remaining panels, for each of the simulations we compare outflow rates at three scales: $r=0.25\times R_{\rm 200c}$, $r=1.00\times R_{\rm 200c}$, and $r=2.50\times R_{\rm 200c}$. Firstly, focusing on the left-hand panels which compare the simulations at the ISM scale, we note that at both redshifts, below \logmhalo$\approx 12$, total outflow rates in \tng\ are actually higher than in \simba, in contrast to the mass loading factors which are higher in \simba. This can be explained by the slightly lower normalisation of the $M_{\star}-{\rm sSFR}$ relation in \simba\ compared to \tng\ (see Fig. \ref{fig:results2:mstar_etaism_z2} for sSFR measurements at $z\approx2$ and \citealt{Dave2019} for measurements at $z\approx0$) -- meaning that galaxies in \simba\ eject more gas {\it per unit SFR} compared to \tng\ galaxies, but have lower absolute total outflow rates. This highlights the fact that if a galaxy is gas poor, the amount of gas available for removal is proportionally reduced. 

It is when we analyse the behaviour of outflow rates {\it as a function of scale} as per the right-hand panels in Fig. \ref{fig:results2:m200_outflow_scales_z2} and \ref{fig:results2:m200_outflow_scales_z0}, that dramatic differences between gas flows in the simulations begin to emerge. We first focus on the $z\approx2$ case, as presented in Fig. \ref{fig:results2:m200_outflow_scales_z2}. Below \logmhalo$\approx12$ in \eagle\ (where stellar feedback is dominant in driving outflows), mass outflow rates at the ISM ($0.25\times R_{\rm 200c}$), halo ($1\times R_{\rm 200c}$), and IGM ($2.5\times R_{\rm 200c}$) scales remain very similar. This indicates that the galactic winds in \eagle\ can travel to at least $2.5\times R_{\rm 200c}$ at $z\approx2$ before stalling and falling back. 

Unlike the picture in \eagle, below \logmhalo$\approx12$ outflow rates {\it decrease} with increasing scale in \tng. At \logmhalo$\approx11.5-12$ in \tng, outflow rates at $r=2.5\times R_{\rm 200c}$ are just $10\%$ of those measured at $r=0.25\times R_{\rm 200c}$. We do note, however, that this stalling is less obvious in lower-mass systems with \logmhalo$\approx11$ -- at this mass, average outflow rates at $R_{\rm 200c}$ are $\approx50\%$ of that measured at the ISM-scale. In general, the stalling of SF-driven outflows in \tng\ is related to the velocities imparted to wind particles at injection, which scale with $\sigma_{\rm DM}$, which is almost always below the escape velocity of the halo \citep{Nelson2019}. This means that SF-driven outflows in \tng\ are ultimately confined to recycle within the halo, except for perhaps in the context of very low-mass haloes ($\lesssim10^{10.5}\, {\rm M}_{\odot}$). In \simba, the picture for SF-driven outflows is somewhere between \eagle\ and \tng. At low halo masses ($M_{\rm 200c}\approx10^{10.5}\, {\rm M}_{\odot}$) outflows are able to reach a similar scale as predicted in \eagle\ ($2-3\times R_{\rm 200c}$), while towards larger masses ($M_{\rm 200c}\approx10^{12}\, {\rm M}_{\odot}$) outflows largely stall before reaching $R_{\rm 200c}$. This result is again a direct consequence of the assumed wind velocities at injection, which in \simba\ are taken from \citet{Muratov2015}.

Above \logmhalo$\approx12.5$ at $z\approx2$, where AGN feedback begins to dominate the driving of outflows, the behaviour between the simulations is also different. In \eagle, the outflow rate at the ISM and halo scales remain similar in this mass range, but drop dramatically at the $2.5\times R_{\rm 200c}$ scale, with $\dot{M}_{\rm out}$ at $2.5\times R_{\rm 200c}$ approximately one tenth of the value at $R_{\rm 200c}$ in halos with mass \logmhalo$\approx13$. This indicates that AGN-driven outflows in \eagle\ tend to stall near or just outside the virial radius at $z\approx2$, and seldom reach larger scales. \tng\ shows qualitatively similar behaviour, but a less dramatic drop off in outflow rate with increasing scale. \simba\ actually shows a monotonically increasing outflow rate with increasing scale at \logmhalo$\approx11.5-13$, and a constant outflow rate from $R_{\rm 200c}$ to $2.5\times R_{\rm 200c}$ at higher halo masses.

This indicates that the outflows driven by AGN in \tng\ and \simba\ are more powerful in terms of spatial reach, likely related to the fact that in both of these models, AGN feedback is implemented with a significant kinetic component rather than just thermally as in the case of \eagle. 

Moving to the $z\approx0$ Universe in Fig. \ref{fig:results2:m200_outflow_scales_z0}, we note that the qualitative differences between simulations remain similar to the $z\approx2$ picture at the ISM scale, however the disparities between how the outflows propagate across scales become even more pronounced. In the left-hand panel, we can see that the normalisation of gas flow rates relative to halo mass are much lower than at $z\approx2$, with longer dynamical time-scales and lower gas densities at this epoch. One exception to this is the strength of AGN-driven outflows in \simba\ -- which are much stronger at $z\approx0$, and also become efficient at lower halo masses. This is likely a result of jet-mode feedback switching on more frequently at low redshift due to the requirement of low Eddington ratios (typical Eddington ratios decline at low redshift; see \citealt{Thomas2019}, Fig. 5). 

\begin{figure*}
\includegraphics[width=\textwidth]{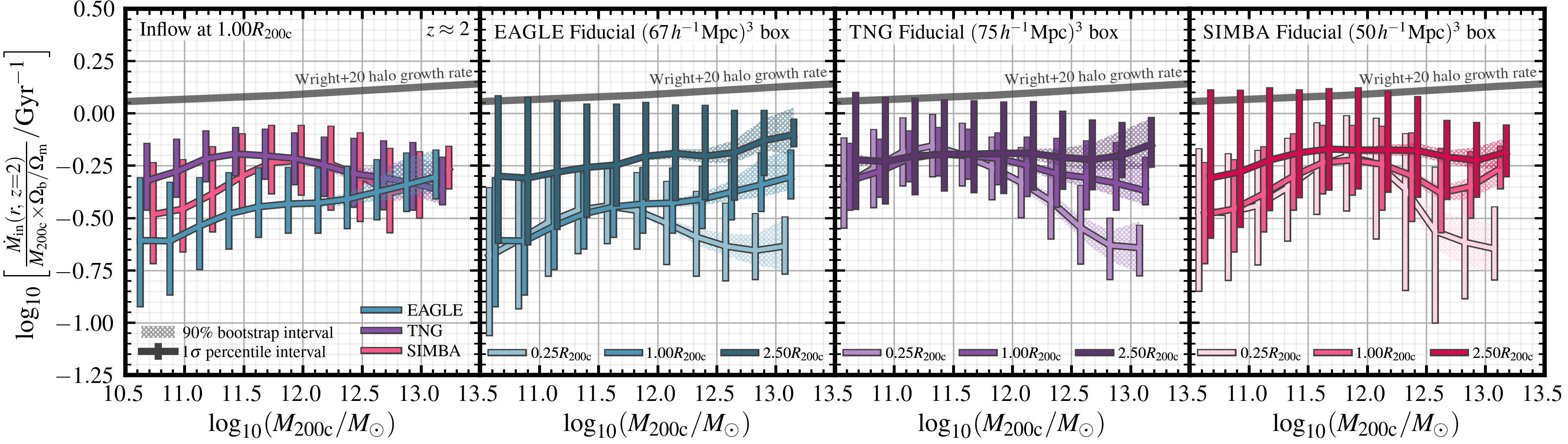}
\caption[]{Mass inflow rates as a function of $M_{\rm 200c}$ at $z\approx2$ in \eagle\ (blue), \tng\ (purple) and \simba\ (pink). Inflow rates are normalised by $f_{\rm b}\times M_{\rm 200c}$ in each case to represent gas inflow ``efficiency'' across different mass scales. {\it Left panel}: Inflow efficiency at the halo scale, compared between the simulations. {\it Panels 2-4}: Inflow efficiencies at different scales in \eagle, \tng, and \simba\ respectively. In each of these panels, ISM-scale inflows are plotted with lighter colours, halo-scale inflows are plotted in medium colours, and IGM-scale inflows are plotted in darker colours. We also include measurements from \citet{Wright2020} indicating the gas accretion rate expected if gas inflows perfectly traced the infall of DM in grey. In all panels, error-bars correspond to the $16^{\rm th} - 84^{\rm th}$ percentile range in inflow rates at a given mass, and hatched regions correspond to the bootstrap-generated $90\%$ confidence interval on the medians at a given mass. All simulations predict that AGN feedback has a strong preventative impact on gas accretion at the ISM scale, however there is considerable variability between the preventative impact of stellar feedback.  }
\label{fig:results2:m200_inflow_scales_z2}
\end{figure*}

\begin{figure*}
\includegraphics[width=\textwidth]{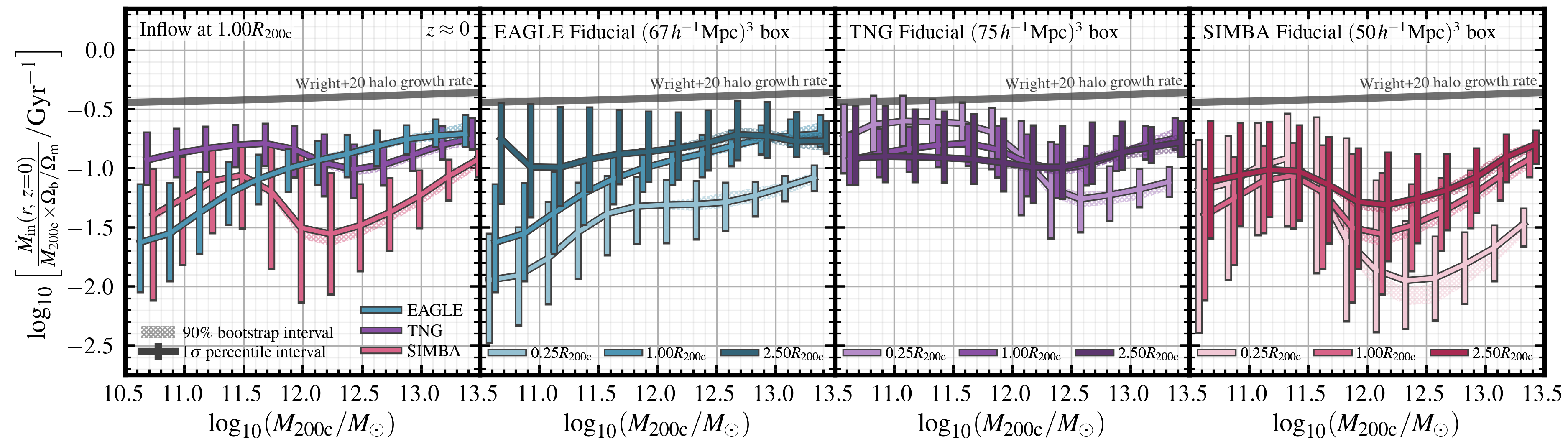}
\caption[]{Mass inflow rates as a function of $M_{\rm 200c}$ at $z\approx0$ in \eagle\ (blue), \tng\ (purple) and \simba\ (pink). Inflow rates are normalised by $f_{\rm b}\times M_{\rm 200c}$ in each case to represent gas inflow ``efficiency'' across different mass scales. {\it Left panel}: Inflow efficiency at the halo scale, compared between the simulations. {\it Panels 2-4}: Inflow efficiencies at different scales in \eagle, \tng, and \simba\ respectively. In each of these panels, ISM-scale inflows are plotted with lighter colours, halo-scale inflows are plotted in medium colours, and IGM-scale inflows are plotted in darker colours. We also include measurements from \citet{Wright2020} indicating the gas accretion rate expected if gas inflows perfectly traced the infall of DM in grey. In all panels, error-bars correspond to the $16^{\rm th} - 84^{\rm th}$ percentile range in inflow rates at a given mass, and hatched regions correspond to the bootstrap-generated $90\%$ confidence interval on the medians at a given mass. The prevention of gas inflow at the halo-scale due to stellar feedback is strongest in \eagle, followed by \simba, then \tng. The prevention of gas inflow at the halo-scale due to AGN feedback is strongest in \simba, followed by \tng, then \eagle. }
\label{fig:results2:m200_inflow_scales_z0}
\end{figure*}

One very interesting feature at $z\approx0$ in the \eagle\ simulations is the apparent entrainment of outflowing  gas with increasing scale. We find that at \logmhalo$\approx11$, the mass outflow rate at $2.5\times R_{\rm 200c}$ is over 10 times higher than that at $0.25\times R_{\rm 200c}$ in \eagle\ at $z\approx0$. This means that the stellar feedback driven outflows in \eagle\ are not only able to reach scales beyond the halo, but also entrain mass along the way. \citet{Mitchell2020a} have previously investigated this effect, finding that the out-flowing gas is significantly over-pressurised relative to the ambient CGM at $z=0$. This causes the  tenuous material in the CGM to be swept up with the bulk flow.  

As demonstrated in Fig. \ref{fig:results2:m200_outflow_scales_z2}, in the regime dominated by stellar feedback, $M_{\rm halo}\lesssim 10^{12}{\rm M}_{\odot}$, we find that in \eagle\ there is minimal mass entrainment in outflows at $z\approx 2$ compared to $z\approx0$ (with the exception of haloes below $10^{11}{\rm M}_{\odot}$, where outflows at $2.5\times R_{\rm 200c}$ are enhanced relative to the ISM and halo scale by a factor of a few). This means that while the outflows do not significantly stall for halo masses of \logmhalo$\approx11-11.5$ at $z\approx2$, we do not find any strong evidence for an increase in mass outflow rates at larger scales as is predicted at later times in \eagle. 

This result can be interpreted with the aid of the density and temperature profiles of haloes that we include in Appendices \ref{sec:apdx:densityprofs} and \ref{sec:apdx:tempprofs}. Comparing the radial density profiles of gas in \logmhalo$\approx11$ at $z\approx2$ and $z\approx0$ in Fig. \ref{fig:apdx:r200_densityprof_z2} and \ref{fig:apdx:r200_densityprof_z0}, we find that in all simulations, gas densities in the CGM universally decrease from $z\approx2$ to $z\approx0$. In \tng\ and \simba, the difference between gas densities at $z\approx2$ and $z\approx0$ is $\approx 1.3-1.5\, {\rm dex}$ at $r=0.50\times R_{\rm 200c}$. This is not an unexpected result -- this decrease in density roughly corresponds to the cubed change in scale factor between these redshifts. In the case of \eagle, however, the decrease in CGM gas density is much more dramatic, closer to $\approx 2-2.5\, {\rm dex}$ from $z\approx2$ to $z\approx0$. This means that at $z\approx0$, outflows are over-pressurised in a relative sense to the CGM (as per the findings in \citealt{Mitchell2020a}). Comparing this picture to $z\approx2$, where CGM densities between all simulations are similar, the lack of mass entrainment at $z\approx2$ in \eagle\ is likely the result of a much denser CGM at this redshift compared to $z\approx 0$, meaning that outflows are not over-pressurised relative to the surrounding gas, and not able to sweep up the surrounding material with the bulk flow. 

An additional factor which is helpful in interpreting different levels of gas entrainment in outflows is the temperature of the gas in the CGM, as presented in Appendix \ref{sec:apdx:tempprofs}. Inspecting the temperature profiles of haloes at $z\approx2$ (Fig. \ref{fig:apdx:r200_Tprof_z2}), in the \logmhalo$\approx11$ and \logmhalo$\approx12$ bins, median CGM gas temperatures in \eagle\ are substantially lower than $T_{\rm vir}$ (estimated from \citealt{VandeVoort2017_chapter}); particularly where $r/R_{\rm 200c}\lesssim 0.75$ (the ``inner'' CGM). This aligns with the commonly accepted picture that ``cold-mode'' accretion -- cool, filamentary inflow from the cosmic web -- is dominant at higher redshift and lower halo masses, and that these inflows can constitute a significant portion of the CGM without being shock-heated (e.g. \citealt{Keres2005}). Comparatively, at $z\approx0$ in \eagle, gas temperatures in the CGM are significantly higher -- within $0.2-0.3\, {\rm dex}$ of $T_{\rm vir}$ at \logmhalo$\approx12$. The low covering fractions typical of gas inflows at high redshift (e.g. \citealt{Wright2021}) mean that feedback-driven outflows are less likely to encounter (and thus entrain) such gas in the CGM. 

As discussed above, we find that in \tng, outflow rates monotonically {\it decrease} with increasing scale below \logmhalo$\approx12$ at both $z\approx0$ and $z\approx2$, indicating stalling of outflows as they propagate to larger distances. Interestingly, at $z\approx2$, Fig. \ref{fig:apdx:r200_Tprof_z2} shows that for \tng\ haloes in bins \logmhalo$\approx11$ and \logmhalo$\approx12$, gas in the inner CGM is significantly hotter than in \eagle. The inner CGM likely corresponds to the scale at which much of the stellar feedback energy is thermalised. Unlike \eagle, we find that gas densities in the CGM remain relatively high in this mass range at $z\approx0$ in \tng, which in addition to the relatively small thermal component of stellar feedback in \tng\ ($10\%$ of feedback energy), contributes to the outflowing material being under-pressurised relative to the surrounding medium, and thus unable to reach larger scales or entrain mass. 

Considering larger haloes at $z\approx0$, an interesting feature in \simba\ is the efficient entrainment of AGN-driven gas outflows above \logmhalo$\approx12$. At \logmhalo$\approx12.5$, the gas outflow rate at $2.5\times R_{\rm 200c}$ is roughly 5-10 times the ouflow rate at $r=0.25\times R_{\rm 200c}$ -- an effect not seen in \eagle, nor \tng. At $z\approx0$, the ambient density of the CGM at this mass scale is quite low (as  demonstrated in Fig. \ref{fig:results1:m200_scalings_z0} and \ref{fig:apdx:r200_densityprof_z0}). This, together with the direct heating of gas elements surrounding AGN, means that the AGN-driven outflows are over-pressurised relative to the surrounding medium, and thus able entrain tenuous gas in both the CGM, and the region surrounding the halo. 

\subsection{Preventative feedback: gas \textit{\textbf{inflow}} rates as a function of scale}\label{sec:results:2:inflows}

In this section, we build on the results of \S\ref{sec:results:2:outflows} to study the gas {\it inflow} rates onto haloes as a function on halo mass, scale, and redshift in the \eagle, \tng, and \simba\ simulations. This helps us understand feedback not only from an ``ejective'' perspective, but also a ``preventative'' perspective; where feedback actively prevents the introduction of baryons to haloes, in addition to directly removing gas via outflows. 
This is very important for semi-analytic and semi-empirical models of galaxy formation, which typically assume that the amount of gas accretion traces the growth of the DM halo (\citealt{Moster2018,Behroozi2019,Pandya2020,Wright2020}).

In Fig. \ref{fig:results2:m200_inflow_scales_z2} and \ref{fig:results2:m200_inflow_scales_z0}, we show the relationship between inflow rate and halo mass in each of the simulations at $z\approx2$ and $z\approx0$. In the left hand panel, we plot inflow rates (normalised by $M_{\rm 200c}\times\ \Omega_{\rm b}/\Omega_{\rm m}$; `` inflow efficiency'') at the $R_{\rm 200c}$ halo boundary as a function of $M_{\rm 200c}$ for \eagle, \tng\ and \simba, compared with the ``expected'' gas accretion rates from DM accretion rates, computed by multiplying the DM accretion rate times the cosmological baryon fraction (grey lines, \citealt{Wright2020}). The ``inflow efficiency'' measurements plotted here correspond to the inverse of the time-scale that would be required for a gas mass of $M_{\rm 200c}\times\ \Omega_{\rm b}/\Omega_{\rm m}$ to be accreted at this scale. In the $2^{\rm nd}-4^{\rm th}$ columns, we compare inflow rates within the same simulations between three scales: $0.25\times R_{\rm 200c}$ (within the ISM), $R_{\rm 200c}$, and at $2.5\times R_{\rm 200c}$. This comparison allows us to see how inflow rates change from cosmological (IGM) scales down to the ISM scale in each of the simulations. 

First focusing on the left-hand panel for the $z\approx0$ case in Fig. \ref{fig:results2:m200_inflow_scales_z0}, we remark that the respective gas accretion rates between the simulations at $R_{\rm 200c}$ follow a similar functional form to the halo-wide baryon fractions presented in Fig. \ref{fig:results1:m200_scalings_z0}. In the stellar feedback-dominated regime, \logmhalo$\lesssim 11.5$, the lowest specific accretion rates are seen in \eagle, where gas inflow rates are just $\approx10\%$ of the expectation based on the halo growth rate, due to the far-reaching influence of stellar feedback (as directly demonstrated in \S\ref{sec:results:2:outflows}). Accretion rates onto haloes of the same mass are much higher in \tng, with gas accretion rates $\approx50\%$ of the expectation based on the DM halo growth rate. Halo-scale gas accretion in \tng\ remains fairly high due to the reduced efficiency of outflows beyond $R_{\rm 200c}$ in this mass range, as was demonstrated in Fig. \ref{fig:results2:m200_outflow_scales_z2} and \ref{fig:results2:m200_outflow_scales_z0}. The predictions from \simba\ regarding gas inflow rates in the stellar-feedback dominated regime fall between those of \eagle\ and \tng -- where stellar feedback has a stronger preventative effect than in \tng at the halo-scale, but not as dramatic as in the case of \eagle. 

In the AGN-dominated regime, \logmhalo$\gtrsim 12.5$, the simulations predict very different levels of preventative feedback. This is most dramatic in \simba\ at \logmhalo$\approx12-12.5$, where the amount of gas inflow to haloes is less than $10\%$ of the expectation based on the total halo growth rate times the cosmological baryon fraction. The halo-scale gas inflow efficiency in \eagle\ continues to rise with halo mass, indicating that AGN feedback plays little role in preventing gas inflow at this scale (see \citealt{Wright2020}, who report a decrease in halo gas accretion of only $\approx20\%$ compared to the \eagle\ run with no AGN feedback). In \tng, we note a small drop in inflow efficiency due to AGN, where at \logmhalo$\approx12.5$, the inflow efficiency drops to $\approx25\%$ of that expected from the total halo growth rate -- less dramatic than seen in \simba, but about twice as much suppression as predicted by \eagle. In all simulations, approaching halo masses of \logmhalo$\approx13.5$, the ability of AGN to suppress halo-scale gas accretion decreases; related to the reduced influence of outflows at larger scales which are unable to overcome the deeper potential well (see the decreases in outflow rates at $2.5\times R_{\rm 200c}$ at this mass scale in Fig. \ref{fig:results2:m200_outflow_scales_z2} and \ref{fig:results2:m200_outflow_scales_z0}, with the exception of \simba\ at $z\approx0$). 

In the three right-hand panels of each Figure, we explore how inflow rates in \eagle, \tng, and \simba\ vary across scales. In \eagle, at $z\approx0$, we clearly see that the rate of gas inflow decreases with radius across all halo masses. Below \logmhalo$\approx 11.5$, the biggest drop in inflow efficiency occurs between $r=2.5\times R_{\rm 200c}$ and $r=1\times R_{\rm 200c}$, indicating that this is where a large portion of the cosmological inflow begins to encounter the outflowing and/or recycling material, in agreement with our findings discussed in \S\ref{sec:results:2:outflows} regarding the scale at which outflows stall in \eagle. At \logmhalo$\gtrsim12$, we note that while cosmological inflow is not greatly decreased at the halo-scale in \eagle, the rate of accretion is significantly suppressed at $r=0.25R_{\rm 200c}$ (as also reported in \citealt{Correa2018a,Correa2018b}). This agrees with our mass-loading results, which suggest that the strength of AGN-driven outflows at the scale of the ISM are similar between \eagle\ and \tng, but that they stall at smaller radius in \eagle. 

Inspecting inflow rates across different scales in \tng, we find that for haloes with \logmhalo$\lesssim 11.5$, inflow rates at the ISM scale ($0.25\times R_{\rm 200c}$) are similar to inflow rates at the halo boundary at $z\approx2$; but at $z\approx0$, they are actually twice as high at the scale of the ISM as those recorded at the scale of the halo. With the relatively dense CGM in \tng\ in this mass range (see Fig. \ref{fig:apdx:r200_densityprof_z0}, partly contributed to by the build-up of ejected, metal enriched gas in this region which does not leave the halo), we argue that the ISM-scale inflow rates are higher than halo-scale inflow rates due to a combination of efficient recycling of cold gas, and efficient metal cooling of warm CGM gas at this scale. Above \logmhalo$\approx 12.5$, we find the opposite effect -- ISM-scale inflow rates are suppressed relative to halo-scale inflow rates (as is true for all simulations in this mass regime).  

Lastly, focusing on inflow rates in \simba, we find that below \logmhalo $\lesssim 11.5$, accretion rates are suppressed equally at $r=0.25R_{\rm 200c}$, $R_{\rm 200c}$, and $2.5R_{\rm 200c}$. Based on Fig. \ref{fig:results2:m200_outflow_scales_z2} and \ref{fig:results2:m200_outflow_scales_z0}, this aligns with a scenario where the spatial range of stellar feedback sits between that of \eagle\ (far-reaching, beyond the halo) and \tng\ (within the inner CGM). For \logmhalo $\gtrsim 12$, \simba\ predicts a steadily increasing inflow prevention effect moving inwards from $r=2.5\times R_{\rm 200c}$ to $r=0.25\times R_{\rm 200c}$. Even though the preventative effect is greatest at $0.25\times R_{\rm 200c}$ (with gas inflow rates at just 1-5$\%$ of those expected based on the halo growth rate at $10^{12.5} {\rm M}_{\odot}$), \simba\ predicts that even gas at $2.5\times R_{\rm 200c}$ experiences significant preventative feedback due to AGN. While we do not present the results here, we note that in the \simba\ run with no jet-mode feedback, the preventative effect is greatly reduced -- indicating that it is the low-accretion rate, high kinetic energy jet-mode AGN feedback in \simba\ that is responsible for the strong suppression of gas inflow. 

To summarise -- in \eagle, \tng\ and \simba, the level of inflow prevention at a given scale can be clearly linked with the presence, or lack thereof, of feedback-driven outflows at the same scale. For \logmhalo$\lesssim 11.5$, \eagle\ and \simba\ predict that stellar feedback produces a strong preventative effect on halo-scale inflow rates, while in \tng, halo-scale gas inflow is suppressed by a smaller amount. For halo masses  \logmhalo$\gtrsim 12$, \simba\ predicts very strong AGN-driven suppression in halo-scale gas inflow rates, relative to a more mild preventative effect in \tng, and very minimal preventative feedback at the halo scale in \eagle. 

\section{Discussion}\label{sec:discussion}
 The \eagle, \tng, and \simba\ simulations have all been largely successful in producing a reasonable match to the observed galaxy stellar mass function at $z\approx0$, albeit with careful tuning required to calibrate their respective sub-grid feedback models that cannot be implemented \textit{ab initio}. We have demonstrated that different implementations of feedback processes -- and thus, the manner in which the baryon cycle operates over cosmic time -- can be degenerate in producing galaxies at $z\approx0$ with very similar stellar masses and star formation rates. In \S\ref{sec:discussion:haloproperties}, we use our measurements of gas flows rates across the simulations in \S\ref{sec:results:2} to make the causal link between the baryon cycle and static halo properties. In \S\ref{sec:discussion:prevlit}, we compare our results to previous literature investigating the baryon cycle in simulated galaxies. In \S\ref{sec:discussion:observations} we outline a number of promising observational avenues by which the degeneracy between different feedback models could be broken. Lastly, in \S\ref{sec:discussion:resolution}, we discuss how the finite resolution of the simulations used in this study may influence our results.

\subsection{The influence of gas flows on halo baryon content}\label{sec:discussion:haloproperties}

\begin{figure*}
\centering
\includegraphics[width=0.98\textwidth]{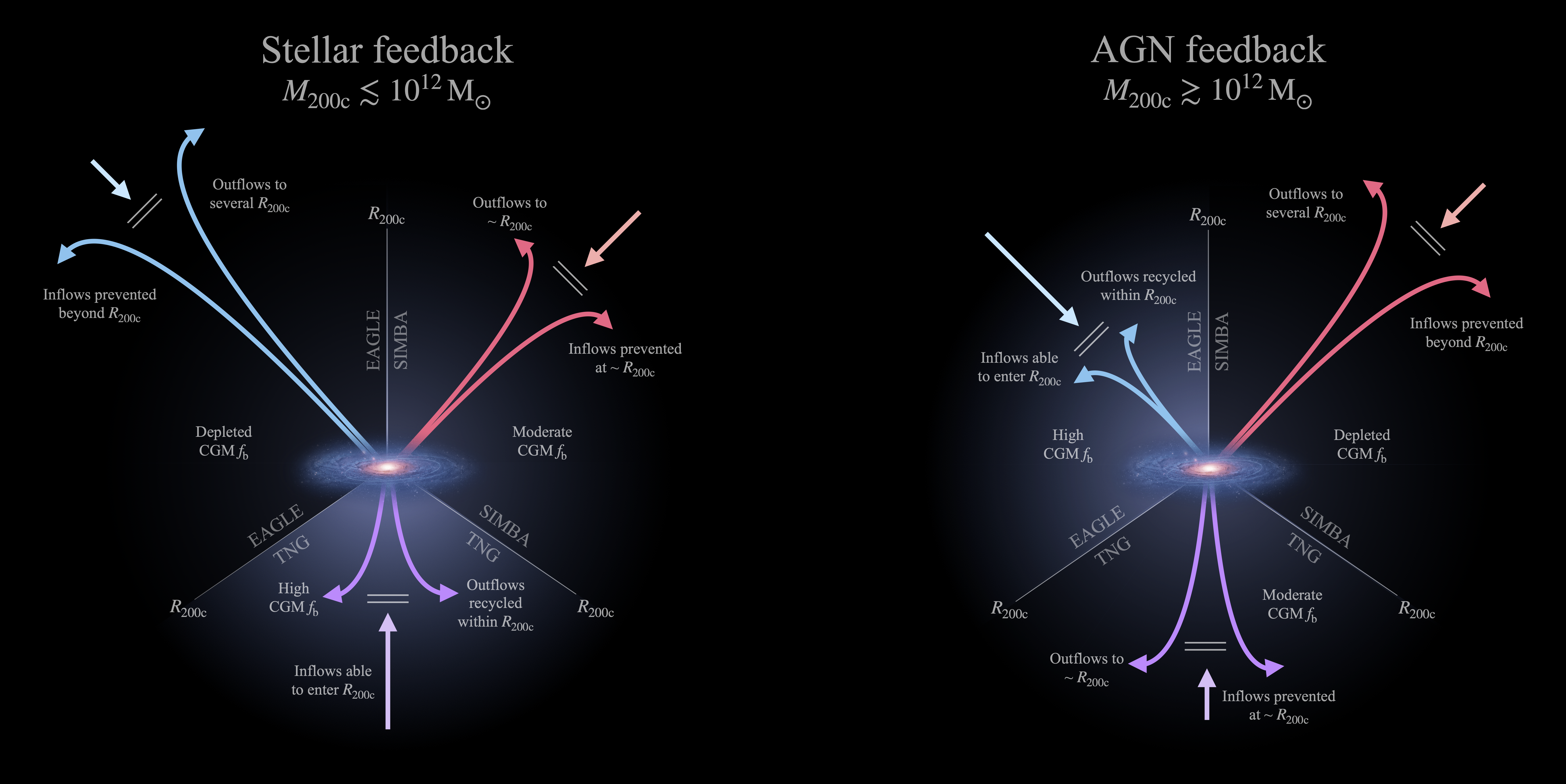}
\caption[]{Schematic diagram summarising the different gas flow paradigms found in \eagle, \tng\ and \simba, split into low mass haloes where stellar feedback dominates (left), and higher mass haloes where AGN feedback dominates (right). }
\label{fig:discussion:flowsketch}
\end{figure*}

In Figure \ref{fig:discussion:flowsketch}, we summarise the main findings of this study in visual form -- indicating the scale reached by feedback-driven outflows in each of the simulations and the consequent gas content of the CGM in each of the simulations. We split this diagram into the low mass ($M_{\rm 200c}\lesssim10^{12}{\rm M}_{\odot}$) and high mass ($M_{\rm 200c}\gtrsim10^{12}{\rm M}_{\odot}$) regimes, where stellar and AGN feedback respectively dominate. Our findings in \S\ref{sec:results:2} aid in the interpretation of our findings in \S\ref{sec:results:1} -- in particular Fig. \ref{fig:results1:m200_scalings_z0} -- where we compare the total baryon content of haloes in each simulation, and its breakdown into stellar and gaseous components. Firstly, we discuss the low-mass regime, \logmhalo$\lesssim12$, where stellar feedback dominates the behaviour of the baryon cycle. In \eagle, the efficient removal of gas to large scales -- where outflows reach several times the virial radius at both $z\approx2$ and $z\approx0$ -- can be clearly linked with the reduced halo gas content below \logmhalo$\approx11.5$ in Fig. \ref{fig:results1:m200_scalings_z0}. This is maintained by the reduced halo gas inflow rates as a consequence of direct prevention of cosmological gas accretion, as shown in \S\ref{sec:results:2:inflows}. 

In contrast, in \tng, stellar feedback driven outflows typically stall before reaching $R_{\rm 200c}$. This deposits a significant amount of enriched gas in the CGM that can subsequently be efficiently recycled into the ISM, resulting in increased ISM-scale inflow rates relative to even halo-scale inflow in \tng\ at $z\approx0$. This efficient intra-halo recycling scenario means that in low-mass haloes, the baryon content of the CGM in \tng\ is much higher than that recorded in \eagle\ at $z\approx0$. \simba\ predicts a picture in between that of \eagle\ and \tng, where some gas is fully removed from the halo via stellar feedback (particularly below \logmhalo$\approx11$), and some portion of these outflows stall within the CGM. This leads to halo-wide baryon fractions and CGM gas content between that of \eagle\ and \tng\ at $z\approx0$. 

Next, we discuss the high-mass regime -- \logmhalo$\gtrsim12$ -- where AGN feedback becomes dominant in these simulations, and plays a driving role in the baryon cycle. In \eagle, we find that the purely thermal implementation of AGN feedback can remove gas from the ISM, but is seldom powerful enough to remove gas to large scales and significantly influence cosmological inflow. This is likely a consequence of rapid thermal dissipation as the gas travels through the CGM, meaning that the outflows stall before travelling far beyond the halo. Even so, this thermal energy (in combination with the transition to a hot, virialised halo atmosphere -- e.g. \citealt{Correa2018a,Correa2018b}) is sufficient to prevent cosmological inflow from entering the ISM (see Fig. \ref{fig:results2:m200_inflow_scales_z2} and \ref{fig:results2:m200_inflow_scales_z0}), and ultimately curtail star formation rates in high mass galaxies. Since little gas is expelled from the halo and the preventative impact of AGN works within the CGM, this leads to the result shown in Fig. \ref{fig:results1:m200_scalings_z0} where galaxy baryon fractions monotonically increase with halo mass. 

In \S\ref{sec:results:2}, we show in Fig. \ref{fig:results2:m200_outflow_scales_z2} and \ref{fig:results2:m200_outflow_scales_z0} that AGN-driven outflows in \tng\ and \simba\ reach much larger scales before stalling, presumably because of the kinetic implementation of feedback in the "jet mode" in both models. At $z\approx0$ we show that for haloes with \logmhalo$\approx12.5$, gas outflow rates at $2.5\times R_{\rm 200c}$ are just as high as they are at $R_{\rm 200c}$ in \tng, and even higher in the case of \simba. Unlike in \eagle\ (where the preventative effect of AGN is confined to within the CGM), these outflows also act to prevent cosmological gas accretion at the halo-scale. Together, particularly in the case of \simba, this causes the downturn in halo baryon fractions at intermediate halo mass which we demonstrate in Fig. \ref{fig:results1:m200_scalings_z0}.

\subsection{Comparison to previous literature}\label{sec:discussion:prevlit}
In this section, we outline how our results compare with previous theoretical studies relating to the baryon cycle of galaxies. In particular, we focus on comparing how outflow and inflow rates change with scale relative to the simulations we study here. 

As outlined in \S\ref{sec:methods:flows} and Appendix \ref{sec:apdx:literature}, we have demonstrated that our methodology for measuring gas flow rates produces results that are consistent with previous studies utilising the \eagle\ simulations \citep{Mitchell2020a} and the \tng\ suite \citep{Nelson2019}. Fig. 14 in \citet{Mitchell2020a} shows the difference between $z\approx2$ outflow mass loading reported by \citet{Nelson2019} in the TNG50 simulations, compared with those measured in \eagle. For galaxies with stellar mass below $\approx10^{10}{\rm M}_{\odot}$, where stellar feedback is efficient, gas outflow rates in TNG50 tend to decrease from smaller ($20\, {\rm kpc}$) to larger ($50\, {\rm kpc}$) radii. Comparatively, in \eagle\, the mass loading is identical at both of these scales. Our work, which uniquely applies a {\it like-for-like methodology} to measure gas flow rates between the simulations, provides a comprehensive explanation for this result -- showing that SF-driven outflows produce a larger-scale impact on the baryon cycle in \eagle\ compared to \tng. 

We can also compare our results with those from zoom-in simulations, where improvements in resolution enable more detailed sub-grid modelling for gas cooling, star formation and feedback within the ISM.\footnote{We note that the FIRE, FIRE-2, NIHAO and \citet{Christensen2016} simulations we discuss do not model AGN feedback.} \citet{Muratov2015} and \citet{AnglesAlcazar2017b} investigate mass loading and baryon cycle predictions from the FIRE simulations, introduced in \citet{Hopkins2014}. FIRE implements stellar feedback by injecting energy, momentum, mass, and metals from stellar radiation pressure, \HII\ photo-ionization and photo-electric heating, Type I and Type II supernovae, and stellar winds -- while not hydrodynamically decoupling or disabling cooling for affected gas elements. The $M_{\star}-\eta$ relations presented in \citet{Muratov2015} and \citet{AnglesAlcazar2017b} (with an Eulerian, shell-based technique and Lagrangian, particle tracking technique respectively) are very similar to those we present for \simba\ in Fig. \ref{fig:results2:mstar_etaism_z2}. This is, perhaps, unsurprising, given that \simba\ uses the FIRE power-law fits as the prescriptive mass loading factor and wind velocity. The  mass loading values at fixed stellar mass in \tng\ and \eagle\ are, thus, lower than the predictions from FIRE (for the mass range considered here, $M_{\star}\gtrsim10^{9}{\rm M}_{\odot}$).

Focusing on an individual system (m12i), \citet{Muratov2015} show that in a halo with $M_{\rm 200c}\approx 10^{12}{\rm M}_{\odot}$ at $z\approx2$, mass loading values at the ISM scale ($0.25\times R_{\rm 200c}$) are stochastic (like the outflows themselves), however typically take a value of $\eta\approx10^{1}$ in episodes of star formation. In this same system, \citet{Muratov2015} show that outflow rates at $R_{\rm 200c}$ reach only $\approx 10-20\%$ of the outflow rate at the ISM in episodes of star formation. This aligns with the predictions we outline from \tng\ and \simba, where SF-driven outflows typically recycle within the CGM at this mass scale. Comparatively, in \eagle, mass flow rates at both radii are very similar at $z\approx2$ (on average). 

\citet{Pandya2021} investigate the mass, momentum, and energy loading in the subsequent FIRE-2 simulations \citep{Hopkins2018}. When using the same radial shell to define the ISM and no minimum outgoing velocity requirement (as per the method in \citealt{Muratov2015}), they find a very similar mass loading scaling with stellar mass. At $z\approx0$ in a halo of mass $M_{\rm 200c}\approx10^{12} {\rm M}_{\odot}$ (m12f), they show that halo-scale gas outflow rates rarely exceed those from the ISM escape -- again, in alignment with the picture painted by \tng\ and \simba\ at this mass. Interestingly, for dwarf haloes with $M_{\rm 200c}\approx10^{10}{\rm M}_{\odot}$, \citet{Pandya2021} find that mass outflow rates at the halo scale nominally {\it exceed} that measured at the ISM scale, and at $M_{\rm 200c}\approx10^{11}{\rm M}_{\odot}$, that outflow rates at the two scales can be roughly equal (if the time delay as the outflow traverses the ISM is taken into account). The former case is similar to the picture painted by \eagle\ at $M_{\rm 200c}\approx10^{11}{\rm M}_{\odot}$, where mass is entrained in outflows through the CGM. This indicates that in FIRE-2, the scale reached by outflows is strongly influenced by halo mass. 

\citet{Christensen2016} use a particle-tracking technique to study gas flows in and around galaxies in a suite of {\sc gasoline}-based simulations \citep{Wadsley2004}. These simulations use a ``blast-wave'' technique (where cooling is temporarily disabled) to distribute energy from stellar feedback, as per \citet{Stinson2006}. As demonstrated in \citet{Mitchell2020a}, \citet{Christensen2016} find lower ISM-scale mass loading factors than those measured in \eagle, and a steeper decline in $\eta$ with mass. Since the ISM-scale mass loading values are lowest in \eagle\ of the 3 simulations we discuss here (see Fig. \ref{fig:results2:mstar_etaism_z2}), this indicates that the mass loading values in \citet{Christensen2016} are also lower than measured in \simba\ and \tng. While direct measurements of halo-scale outflow rates are not presented, they find that a substantial fraction of the gas that leaves the ISM will also leave the halo virial radius. 

\citet{Tollet2019} present an analysis of gas flows in the NIHAO simulations using a Lagrangian particle tracking method \citep{Wang2015}. NIHAO also uses the blast-wave technique to distribute energy from stellar feedback, as per \citet{Stinson2006}. As a function of stellar mass, mass loading values in NIHAO are similar to \tng\ and \simba\ at $M_{\star}\approx10^{9}{\rm M}_{\odot}$ ($\eta\approx10^{1.5}$), however decrease sharply with stellar mass to $\eta\approx10^{-0.5}$ at $M_{\star}\approx10^{10.5}{\rm M}_{\odot}$. At the same stellar mass, predictions from \eagle, \tng, and \simba\ are one full decade higher at $\eta\approx10^{0.5}$. For haloes with $M_{\rm 200c}\lesssim10^{12}{\rm M}_{\odot}$, they show that the majority of gas does not exit $R_{\rm 200c}$ after being ejected from the ISM. 

The total baryon fraction of haloes within $R_{\rm 200c}$ we measure align with the findings presented in \citet{Ayromlou2023}, who also analyse the \eagle, \tng, and \simba\ simulations to show how feedback redistributes the baryons in and around haloes. This is measured in terms of the ``closure radius'' of haloes -- the minimum radius at which the enclosed baryon content reaches the universal value. They find that for low mass haloes where stellar feedback dominates ($M_{\rm 200c}\lesssim10^{11.5}{\rm M}_{\odot}$), the closure radius is the largest for \eagle\ galaxies, followed by \simba\ galaxies, and then \tng\ galaxies. In the case of higher mass haloes where AGN feedback dominates ($M_{\rm 200c}\gtrsim 10^{12}{\rm M}_{\odot}$), the closure radius is the largest for \simba\ galaxies, followed by \tng\ galaxies, and then \eagle\ galaxies. 

Our findings also provide insight into the measurements presented in \citet{Dave2020}, who investigate the global cold gas (\HI\ and \Hmol; decomposed in post-processing) content of galaxies in the same simulations and compare with available observations. At $z\approx0$, \citet{Dave2020} show that the fiducial resolution \eagle\ run under-predicts the \HI\ content of galaxies compared to local ALFALFA observations of the \HI-MF \citep{Jones2018}, as well as the \HI\ fraction of galaxies as a function of $M_{\star}$. Given that the \HI\ identified by this decomposition algorithm tends to trace the gas that is slightly more diffuse and extended than \Hmol\ gas, the low \HI\ content of \eagle\ galaxies could be explained by the stellar feedback scenario we outline above, which evacuates much of the CGM -- and, likely, some of the gas identified as \HI\ which resides the edge of the ISM. We do note, however, that predictions from the recalibrated, higher resolution \eagle\ run and \HI\ observations and are in closer agreement (see also Fig. \ref{fig:apdx:r200_densityprof_z2} and \ref{fig:apdx:r200_densityprof_z0}). 

Our results highlight the fact that differences in baryon cycling can indeed leave signatures even at the scale of the ISM, particularly when considering multi-phase gas. With the resolution and physics improvements in the next generation of cosmological simulations, it will become possible to better resolve the phase structure of the ISM in a self-consistent manner -- and thus, to produce statistical predictions for the spatially resolved properties of the ISM, as well as the influence of gas flows on these distributions. Such advancements will also provide the statistical context necessary to generalise the results of high-resolution zoom simulations (some of which we summarise above), which provide more detailed insight into the link between the ISM and CGM. 

\subsection{Implications for semi-analytic models} \label{sec:discussion:SAMs}
Semi-analytic models (SAMs) have been used extensively as a complement to numerical hydrodynamic simulations to make predictions for how galaxies form and evolve and to interpret observations \citep[for a review see][]{Somerville2015}. Instead of explicitly solving the equations of gravity, hydrodynamics, etc numerically, SAMs instead track bulk flows of mass and metals into and out of reservoirs that typically include gas (and gas phase metals) in the IGM, CGM, and ISM, as well as stars within galaxies. The models are set within cosmological ``merger trees'' which represent the formation of structure and growth of halos via mergers and cosmological accretion with the $\Lambda$CDM paradigm. Phenomenological, parameterized scaling relations describe processes such as the rates of cooling of hot halo gas and its flow into the ISM, conversion of cold ISM gas into stars, ejection of gas by stellar feedback, and accretion onto and feedback from supermassive black holes. These free parameters are typically tuned
to match a set of galaxy observables or quasi-observables, which have traditionally focused on stellar and ISM properties such as stellar mass functions (or stellar mass vs. halo mass relations), cold ISM gas fraction, stellar mass vs. stellar phase metallicity, SFR distributions, etc. Different SAM groups have adopted very different mass loading factors for stellar driven winds (see e.g. the comparison presented in Fig. 9 of \citealt{Mitchell2020a}), in spite of predicting very similar stellar mass functions and stellar-to-halo mass ratios. The results presented here provide important insights for future SAMs.

Some SAMs do not have a direct channel for stellar feedback to eject CGM material from the halo \citep[e.g. Santa Cruz, GALFORM SAMs;][]{Somerville2008, Bower2012,Lacey2016}. Gas and metals can be ejected from the ISM and are either deposited in the CGM or into an ``ejected'' reservoir. Some SAMs compute the total energy budget produced by stellar populations and parameterize the fraction that drives an ISM-scale outflow, and assume that any remaining energy can be used to eject gas from the CGM \citep[e.g.  L-galaxies, SHARK, and related models; ][]{Henriques2015,Lagos2018a}. AGN feedback in some models can also help eject material out of the ISM to the CGM, and potentially from the CGM to the IGM \citep[e.g.][]{Lagos2024}. 

Our study shows that in some hydrodynamical simulations (e.g. \eagle), the ejected mass on CGM scales ($R_{\rm 200c}$) can be higher than that on ISM scales (0.25$R_{\rm 200c}$), indicating that mass is entrained in the winds. This suggests that SAMs could 
include this additional channel for removal of material from the CGM by stellar driven winds, and also provides some guidance for priors on the coupling factors that are currently treated as free parameters. 

In most SAMs, gas and metals in the ejected reservoir can ``re-accrete'' into the CGM on a specified timescale. This timescale, and how it scales with halo mass and time, are uncertain and different choices are adopted in different SAMs. Though not studied explicitly here, our results demonstrate that different hydrodynamic simulations are likely to predict very different timescales for this ``re-accreted'' gas, since they show very different rates of gas accretion at halo scales ($R_{\rm 200c}$). 

SAMs, as well as empirical models \citep{Behroozi2019,Moster2018} typically assume that the mass inflow rate into the CGM is equal to $f_b \dot{M}_{\rm tot}$, where $\dot{M}_{\rm tot}$ is the total rate of mass accretion into the halo\footnote{Most SAMs modulate this to account for a reduced accretion efficiency due to photo-ionization heating by the meta-galactic background, but this typically only affects halos smaller than the mass range studied here ($\lesssim 10^{10} M_{\odot}$).}. As discussed extensively in Section~\ref{sec:discussion:haloproperties}, ``preventative'' feedback (where inflows are considerably smaller than $f_b \dot{M}_{\rm tot}$) can be quite important in simulations, but its strength at different scales varies depending on the subgrid physics. Only a few SAMs have included preventative feedback on halo scales due to stellar feedback processes \citep{Lu2015,Hirschmann2014,Pandya2023}. 

Most SAMs model AGN feedback by allowing thermal energy from a notional radio jet to offset some of the CGM cooling, reducing the rate of cooling from the CGM to the ISM. The AGN is not allowed to eject gas from the CGM to the IGM, or to change the density profile or temperature of the CGM. The Santa Cruz and SHARK SAMs  \citep{Somerville2008,Lagos2024} include mass ejection from the ISM due to radiatively efficient AGN winds. This material is assumed to leave the halo and not to return, due to the high assumed launch velocities in the case of the Santa Cruz SAM, while in SHARK the material may be trapped in the CGM or escape to the IGM depending on the wind energy. In both models, however, this was found to have little impact on the results. It is clear from the results presented here and elsewhere \citep[e.g.][]{Ayromlou2023} that in order to more faithfully depict the baryon cycle in numerical hydrodynamic simulations, SAMs must include recipes for how AGN winds and jets remove gas and metals from the CGM (e.g. \citealt{Lagos2024}), as well as for the associated preventative feedback.  

A promising approach for modeling ejection from the CGM and preventative feedback associated with either stars or an AGN has been proposed by \citet{Carr2023} and \citet{Pandya2023}: in addition to tracking inflows and outflows of mass and metals, these SAMs also include tracking of \emph{energy} sources (e.g. from stellar and AGN feedback) and sinks (e.g. from cooling and turbulent dissipation). Excess energy deposited into the CGM can ``lift'' it out of the halo. An extension of these ideas is presented by Voit et al. (2024a,b, in prep.), in which the CGM is in generalized energy balance within the gravitational potential of the halo, and can expand or contract in response to energy deposition or loss.

\subsection{Identifying observational feedback tests}\label{sec:discussion:observations}

As discussed in \S\ref{sec:introduction}, direct observation and quantification of the gas flows surrounding galaxies is very challenging -- particularly when seeking to compare the baryon cycle at a statistical level. This means that currently, it is very difficult to place constraints on which feedback scenarios are favoured by observations. Here, we briefly outline some avenues against which these models could be tested with current and/or future observations. 

Comparing the average gas density profiles between simulations at $z\approx2$ and $z\approx0$ in Fig. \ref{fig:apdx:r200_densityprof_z2} and \ref{fig:apdx:r200_densityprof_z0}, it is clear that the differences in total halo gas content arising from disparate baryon cycling paradigms are clearest at $z\approx0$, after the variability between models has had adequate time to consolidate in terms of static halo properties. Presently, a combination of X-ray observations and weak lensing studies have been able to place constraints on the gas fraction of high-mass galaxy groups and clusters ($M_{\rm 500c}\gtrsim 10^{13}\, {\rm M}_{\odot}$, e.g. \citealt{Hoekstra2015,Pearson2017,Mulroy2019,Lovisari2020,Akino2022}); which can be used to calibrate the strength of AGN feedback in the next generation of cosmological hydrodynamical models (e.g \citealt{Schaye2023,Kugel2023}). As demonstrated in Figure \ref{fig:results1:m200_scalings_z0}, such observations indicate that the baryon fractions of haloes of mass $M_{\rm 500c}\approx10^{14}\, {\rm M}_{\odot}$ within $R_{\rm 500c}$ are slightly too high in \eagle\ and \tng\ (very close to the universal value), and slightly too low in \simba. 

As demonstrated in Fig. \ref{fig:results1:m200_scalings_z0}, the most significant AGN-driven difference in halo baryon content is induced below the mass scale of clusters, between $M_{\rm 200c}\approx10^{12}-10^{13}\, {\rm M}_{\odot}$. Measurements of the baryon content of haloes in this mass range would provide significant constraining power on the strength of AGN feedback, as well as the mass scale at which SMBH accretion and feedback becomes efficient. Preliminary studies using the eROSITA telescope \citep{Predehl2021} and stacked soft X-ray observations indicate that for galaxies with stellar mass $M_{\star}\approx10^{11}\, {\rm M}_{\odot}$,  forward-modelled X-ray luminosity profiles (up to $\approx R_{\rm 200c}$) from \eagle\ and \tng\ are consistent with observational constraints \citep{Oppenheimer2020,Chadayammuri2022}. Further constraints on the baryon fraction of haloes in this mass range will become possible with observations utilising the Sunyaev Zeldovich (SZ) effect -- for instance, \citet{Bregman2022} demonstrate the feasibility of placing such constraints in a small sample of local $\approx L^{*}$ haloes. \citet{Yang2022} demonstrate that the disparate baryon content of haloes between \tng\ and \simba\ above \logmhalo$\approx12.5$ produces an observationally detectable difference in forward-modelled SZ measurements, provided that an instrument with adequate angular resolution is utilised (such as the upcoming Simons Observatory, \citealt{Ade2019}). In this mass range, studies investigating the \HI\ content of groups using stacking techniques may also prove useful to constraining baryon content (e.g. \citealt{Dev2023}), particularly with radio surveys such as DINGO with the ASKAP telescope \citep{Hotan2021} and MIGHTEE-HI with MeerKAT \citep{Maddox2021}. 

Below mass scales of $M_{\rm 200c}\approx10^{12}\, {\rm M}_{\odot}$, it is very difficult to observationally constrain the baryon content of galaxies and their CGM. In this mass range, the CGM is expected to be multi-phase and anisotropic (e.g. \citealt{McCourt2018,Fielding2020}), constituted of a combination of cool filamentary inflow structures together with enriched recycling gas. Observations have shown that sight-line measurements of CGM metallicity can vary by up to ${\rm 2}\, {\rm dex}$ within the same halo (e.g. \citealt{Lehner2013,Prochaska2017,Zahedy2019}). The resolution of the cosmological hydrodynamical simulations we study here is inadequate to resolve detailed phase structure, and more detailed sub-grid models are also likely required to accurately model the CGM \citep[e.g.][]{Crain2023}. We can, however, make general comments regarding the expected impact of different feedback regimes on the properties of the CGM. 

In the case of \eagle, where the CGM is predicted to be nearly entirely devoid of gas at low halo masses, \citet{Wright2021} show that the integrated metallicity of the CGM is extremely sensitive to the rate of pristine halo-scale gas accretion. This is a result of newly introduced gas being able to efficiently dilute the metallicity of the already small gaseous reservoir, which is only possible due to the previous evacuation of the CGM via far-reaching, strong stellar feedback. Thus, in a model like \eagle, we might expect to find significant amounts of very metal poor to pristine gas in the CGM, which would not be the case in a model like \tng\ where the metal-enriched gas from SF-driven outflows is directly deposited into the CGM. Despite this, we refrain from comparing the models in terms of the distribution of metals in the CGM in this work, as it is non-trivial to disentangle differences in the modelling of metal diffusion (included in \tng, but not in \eagle\ or \simba) from changes induced by differences in gas flows. 

In any case, it is clear that the scale and strength of stellar feedback will leave signatures in the distribution of metals in the CGM. \citet{Peroux2020a} demonstrate that \tng\ and \eagle\ predict similar {\it net} gas flows rates as a function of galacto-centric azimuthal angle for galaxies with $M_{\star}\approx10^{9.5}\, {\rm M}_{\odot}$ and $M_{\star}\approx10^{10.5}\, {\rm M}_{\odot}$. Investigating the {\it scale} of metal-enrichment at different impact parameters, particularly studying how angular {\it and} radial variations would manifest in discrete sight-line observations, could constitute a promising test of model accuracy. This is potentially possible with carefully constructed comparisons with current surveys that probe different phases of the CGM, such as COS-Halos \citep{Tumlinson2013,Werk2014,Prochaska2017} and MEGAFLOW \citep{Schroetter2016,Zabl2019}. 

Alone this line, \citet{DeFelippis2021} compare results from the MEGAFLOW survey with predictions from the TNG50 simulation (several fold higher resolution than the TNG100 simulation we investigate here, but the same feedback physics); indicating that TNG50 can reasonably reproduce observations of the kinematic diversity of strong MgII absorbers (tracing the cool CGM) in the halo mass range $10^{11.5}\, {\rm M}_{\odot}- 10^{12}\, {\rm M}_{\odot}$. \citet{Appleby2021} compare results from \simba\ and the COS-Halos and COS-Dwarfs survey, finding general agreement with observations regarding the predicted abundance of \HI\ and metal absorbers in the CGM of star-forming \simba\ galaxies, but find slight tension in the quenched population. \citet{Appleby2023} also demonstrate the feasibility of using machine learning to infer CGM conditions from absorption line properties, finding that their model trained on \simba\ can accurately predict absorber overdensity, temperature, and metallicities. 

The Habitable Worlds Observatory, a large space based UV-optical-IR mission concept endorsed by the US Astro2020 Decadal report, will provide greatly enhanced spectroscopic sensitivity in the UV relative to the HST Cosmic Origins Spectrograph, enabling significantly larger samples of $z\lesssim 1$ galaxies with absorption line studies of their CGM \citep{Astro2020s}. Furthermore, new ways of probing the CGM continue to be proposed and tested -- for instance using fast radio bursts (FRBs; \citealt{Macquart2020}), and the characterisation of of Ly$\alpha$ haloes in both emission (e.g. \citealt{Lokhorst2019,Augustin2019}) and absorption (i.e. Ly$\alpha$ forest, see \citealt{Tillman2023b,Tillman2023a}). 


Observations of gas flow {\it rates} -- as opposed to static measurements of halo baryon content -- will also prove useful in constraining feedback models in simulations. Different techniques can provide measurements on the multi-phase nature of galactic outflows in individual systems; for instance CO tracing cool molecular outflows (e.g. \citealt{Leroy2015} using ALMA observations), or with metal absorption lines and detailed ionisation modelling (e.g. see discussions in \citealt{Chisholm2016,Veilleux2020}). Combining careful forward-modelling from simulations with observational approaches across wavelength may be able to place constraints on {\it total} gas outflow rates as the sample size of outflow measurements continues to grow.

\subsection{The impact of resolution}\label{sec:discussion:resolution}
It is important to note that the resolution of each of these simulations we discuss is inadequate to resolve the multi-phase structure of the CGM – in particular, any interfaces between cold and hot CGM gas (see e.g. \citealt{McCourt2018,Fielding2020}). \citet{Creasey2011} show that a mass resolution of $<10^6\,{\rm M}_{\odot}$ is desirable to avoid numerical overcooling of accretion shocks onto haloes with SPH – a criterion which is not met in any of the simulations, particularly SIMBA. At even finer resolutions, \citet{Rey2024} show in AMR simulations of a galaxy with mass $M_{\star}$=$10^8\, {\rm M}_{\odot}$ that increasing spatial resolution from $>200\, {\rm pc}$ to $\approx 20\, {\rm pc}$ can lead to two-fold increases in mass loading factors at $r=5\, {\rm kpc}$. They show this is due to the better resolved “hot” outflowing phase ($T\gtrsim10^{5}\, {\rm K} $) being heated to higher temperatures, and staying hot for longer periods. Unfortunately, without any sub-grid refinement schemes, the simulations we analyse here simply cannot simultaneously resolve scales from Mpc down to the $\approx 0.1 - 10 \, {\rm pc}$ required to model cold gas in the ISM or CGM. 

These considerations beg the question as to whether improved resolution could influence the results presented in \S\ref{sec:results:1} \& \S\ref{sec:results:2}. As shown in Table \ref{tab:methods:simtable}, the mass resolution of the \eagle\ and \tng\ simulations are relatively similar, but there is a notable difference in \simba, where the gas element mass is roughly $10$ times higher than the two aforementioned simulations.While we do not show the results here, we have conducted a comparison between each simulation in this paper with a higher resolution counterpart (EAGLE-L25N752-Ref in the case of EAGLE, TNG50 in the case of TNG100, and a higher resolution 25Mpc SIMBA box in the case of SIMBA). 

Upon investigation, we noted that there is very little difference between the propagation of feedback-driven outflows in TNG100 and TNG50. In the case of stellar feedback, this likely due to the fact that feedback affected gas is “decoupled” from the hydrodynamics at injection (directly avoiding any immediate over-cooling), and the mass loading is set based on local DM dispersion as opposed to baryonic properties. At higher halo masses, TNG50 haloes are very slightly more baryon-rich than those in TNG100, however the scale reached by AGN-driven outflows remain identical -- indicating that resolution does not qualitatively influence the conclusions we make in this paper. 

In the case of \eagle\ and \simba, we found slight differences in the propagation of outflows in low mass haloes $\lesssim10^{11.5}{\rm M}_{\odot}$ with enhanced resolution. ISM-scale outflows in both higher resolution \eagle\ and \simba\ runs are enhanced relative to the standard resolution runs, where numerical over-cooling could inhibit the escape of outflows. These slightly enhanced ISM-scale outflow rates relative to the standard resolution runs leads to the deposition of more gas in the CGM, and an associated increase in ISM-scale inflow rates that offset any differences that could otherwise arise in terms of star formation rate.   

Despite these small quantitative differences, we stress that the central findings of our study regarding the differences between sub-grid prescriptions and how this manifests in the larger scale baryon cycle do remain qualitatively sound when comparing with higher-resolution counterparts of each simulation.

\section{Summary}\label{sec:summary}

In this paper, we have conducted a like-for-like comparison of the baryon cycle in three modern cosmological hydrodynamical simulations: \eagle, \tng, and \simba. We use a common methodology to analyse the gas flows in and around central galaxies, from the scale of the ISM to several halo virial radii. While galaxies in these simulations share very similar stellar mass content and star-formation rates at $z\approx0$, our work has highlighted that this agreement is achieved for very different physical reasons.

Concerning haloes with mass $M_{\rm 200c}\lesssim 10^{11.5} {\rm M}_{\odot}$, where stellar feedback is dominant, we summarise our results as follows:

\begin{itemize}
    \item In \eagle, SF-driven outflows are able to reach several times  $R_{\rm 200c}$ before stalling. At $z\approx0$, mass outflow rates increase with scale with the gas surrounding galaxies being very tenuous and under-pressurised relative to the outflows. At $z\approx2$ we do not see mass entrainment in outflows with increasing scale, where the CGM is still relatively dense and more cool/filamentary in nature.  
    \item The extended scale of SF-driven outflows in \eagle\ leads to a significant preventative effect on gas inflow at the halo-scale. 

    Overall, this leads to haloes with a low density CGM and very low baryon fractions (just 10\% of the cosmological value) in this mass range at $z\approx0$. 
    \item In \tng, SF-driven outflows are very strong at the scale of the ISM, however, tend to stall within the CGM before reaching $R_{\rm 200c}$. This is also true in \simba\ at $z\approx2$, however at $z\approx0$, SF-driven outflows in \simba\ can reach a similar scale as in \eagle\ (albeit without mass entrainment). 
    \item In \tng, halo baryon fractions remain above 50\% of the cosmological value in this mass regime at $z\approx0$. SF-driven outflows are recycled within the scale of the halo, with inflow rates at the ISM-scale higher than at the halo-scale due to very efficient cooling within the dense, enriched CGM. 
    \item In \simba, halo baryon fractions in this mass regime are between those measured in \eagle\ and \tng, reflecting moderate preventative feedback on halo-scale gas inflows. 
    
\end{itemize}

\noindent{Furthermore, in relation to haloes with mass $M_{\rm 200c}\gtrsim 10^{12} {\rm M}_{\odot}$, where AGN feedback is dominant:}

\begin{itemize}
\item In \eagle, AGN-driven outflows do not reach far beyond $R_{\rm 200c}$ at either $z\approx0$ nor $z\approx2$. This leads to minimal preventative impact on gas inflow at the halo scale, but significantly prevents gas inflow at the scale of the ISM.  

\item In \tng, AGN-driven outflows are more powerful than in \eagle, with at least some gas in these outflows reaching $2-3\times R_{\rm 200c}$. This leads to a moderate preventative impact on cosmological inflow at the halo-scale, with further inflow prevention within the heated CGM. As a result, there is a significant turnover in halo baryon fractions at $z\approx0$, reaching values as low as 50\% of the cosmological value at $M_{\rm 200c}\approx 10^{12.5}\, {\rm M}_{\odot}$.
\item In \simba, AGN-driven outflows are significantly stronger than in both \eagle\ and \tng, particularly at $z\approx0$. These outflows continually entrain mass from the ISM scale to several times $R_{\rm 200c}$. This causes a very significant drop in halo-scale cosmological gas inflow, and a corresponding drop in halo baryon fractions to as low as 20\% of the universal value between $M_{\rm 200c}=10^{12}{\rm M}_{\odot}-10^{13} {\rm M}_{\odot}$. 
\end{itemize}

This work lays the foundation for developing targeted observational tests that can favour or disfavour certain feedback scenarios, and move towards a more constrained understanding of the role of feedback processes in the baryon cycle. Such an understanding will be imperative to inform the next generation of cosmological simulations, and to maximise their scientific utility. Additionally, a precise understanding of the baryonic effects on the large-scale matter distribution (see \citealt{Delgado2023,Gebhardt2023}) will be imperative in order to extract meaningful constraints from future weak lensing studies with regard to cosmological parameters, cosmic shear, and the growth of cosmic structure.


\section*{Acknowledgements}
RJW acknowledges support from the European Research Council via ERC Consolidator Grant KETJU (no.
818930), the Fulbright Australia Commission for a Postdoctoral Visiting Scholar Grant, and the Australian Research Council Centre of Excellence for All Sky Astrophysics in 3 Dimensions (ASTRO 3D). 

DAA acknowledges support by NSF grants AST-2009687 and AST-2108944, CXO grant TM2-23006X, JWST grant GO-01712.009-A, Simons Foundation Award CCA-1018464, and Cottrell Scholar Award CS-CSA-2023-028 by the Research Corporation for Science Advancement.

For the purpose of open access, the author has applied a Creative Commons Attribution (CC BY) licence  to any Author Accepted Manuscript version arising from this submission.

The authors used the following software tools for the data analysis and visualisation in the paper: 
\begin{itemize}
    \item {\fontfamily{pcr}\selectfont python3} \citep{VanRossum1995}
    \item {\fontfamily{pcr}\selectfont numpy} \citep{Harris2020}
    \item {\fontfamily{pcr}\selectfont scipy} \citep{Virtanen2020}
    \item {\fontfamily{pcr}\selectfont matplotlib} \citep{Hunter2007}

\end{itemize}
\section*{Data Availability}
The \eagle\ (\url{http://icc.dur.ac.uk/Eagle/}), \tng\ (\url{https://www.tng-project.org/}), and \simba\ (\url{http://simba.roe.ac.uk/}) simulation outputs used for our analysis are all publicly available. The gas flow catalogs generated for this work will be shared by the corresponding author upon reasonable request.


\bibliographystyle{mnras}
\bibliography{references.bib}

\appendix

\section{Comparison of gas flow measurements with previous literature}\label{sec:apdx:literature}
In this section, we compare our measurements of gas flow rates with those measured from the same simulations (where applicable) in previous literature. As outlined in \S\ref{sec:methods:flows}, we elect to use an Eulerian or ``instantaneous'' approach to calculating gas flow rates, rather than a Lagrangian particle tracking method. 

\begin{figure}
\includegraphics[width=1\columnwidth]{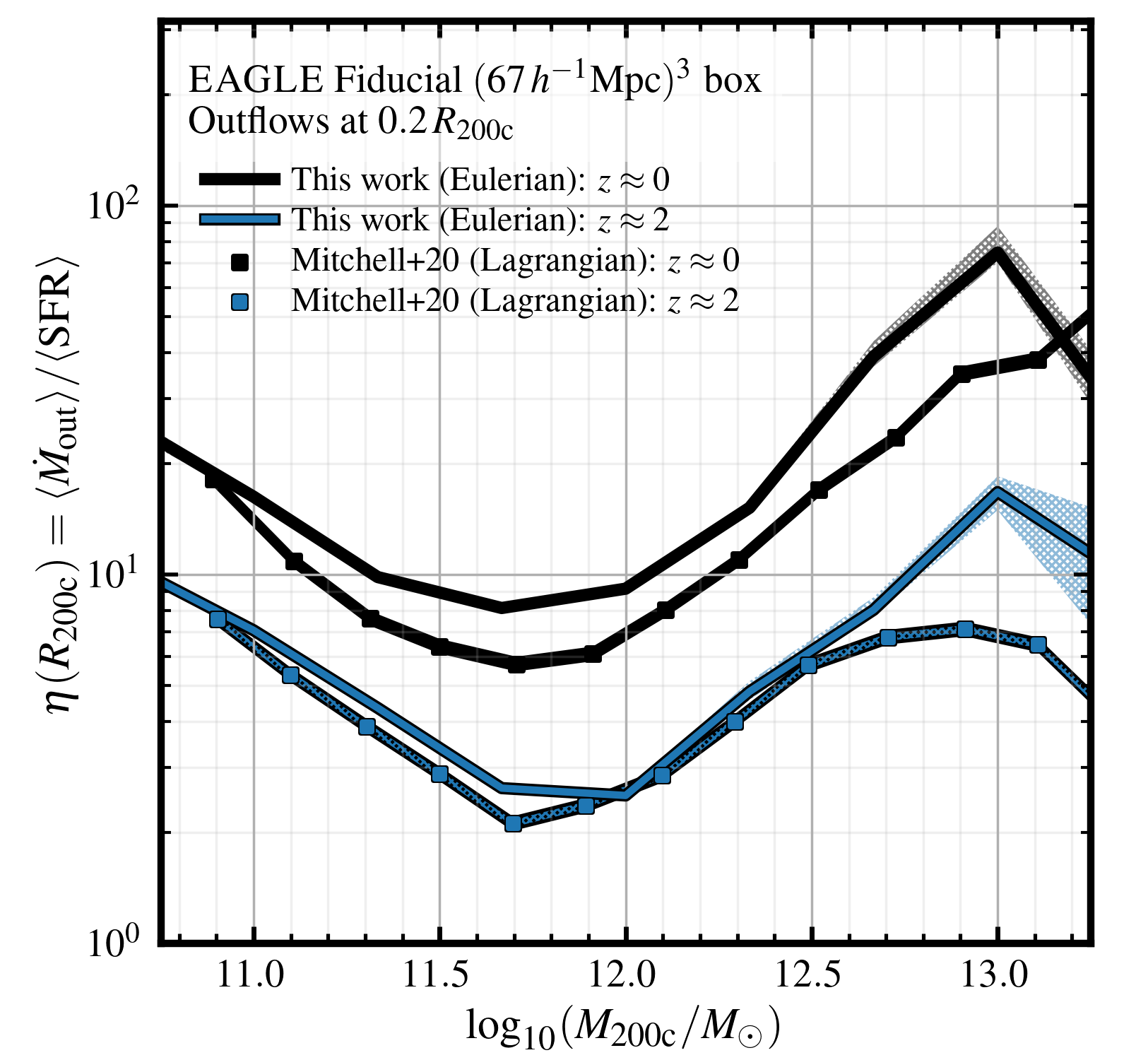}
\caption[]{Comparison of $z\approx2$ (blue) and $z\approx0$ (black) halo-scale mass loading factors ($\eta\equiv\langle\dot{M}_{\rm out}\rangle/\langle\dot{M}_{\star}\rangle$) as a function of $M_{\rm 200c}$ mass measured in this work (solid lines, no shading), and the measurements presented in \citet{Mitchell2020a} (square markers connected with dot-filled solid lines). In our calculations, hatched shaded regions correspond to the bootstrap-generated $90\%$ confidence interval on the medians at a given mass. At both redshifts, our measurements agree well, to within $\approx0.2\,{\rm dex}$.}
\label{fig:apdx:literature_mitchell20}
\end{figure}

Figure \ref{fig:apdx:literature_mitchell20} shows a comparison of $z\approx2$ and $z\approx0$ halo-scale mass loading factors at the ISM scale (here, $0.20\times R_{\rm 200c}$) in \eagle\ as a function of $M_{\rm 200c}$ mass.  While \citet{Mitchell2020a} take a Lagrangian particle-tracking approach to measure gas flow rates, we recover good agreement with their results (to within $\approx0.2\, $dex across the halo mass range). Our instantaneously measured mass loading factors are systematically (very) slightly higher than the \citet{Mitchell2020a} results, likely due to the latter not including the subset of gas that is very quickly recycled within the $\Delta t$ used for the Lagrangian calculation. This offset is highest at $M_{\rm 200c}\approx10^{13}{\rm M}_{\odot}$, indicating that some AGN-driven outflows in \eagle\ may be recycled quite quickly at the ISM scale. 

\begin{figure}
\includegraphics[width=1.02\columnwidth]{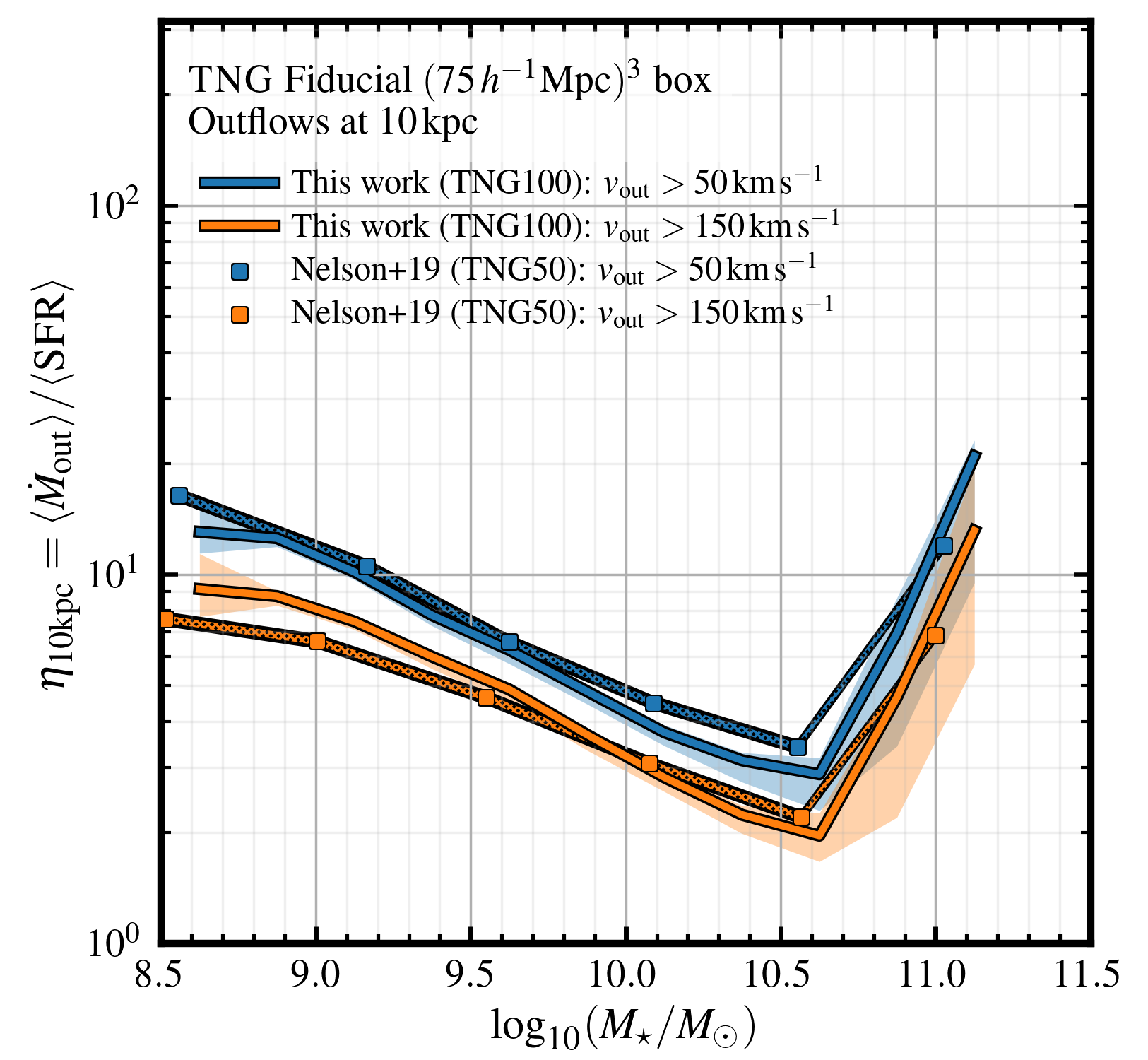}
\caption[]{Comparison of $z\approx2$ mass loading factors ($\eta\equiv\langle\dot{M}_{\rm out}\rangle/\langle\dot{M}_{\star}\rangle$) at $10\, {\rm kpc}$ measured in this work (solid lines, no shading) and those presented in \citet{Nelson2019} (square markers connected with dot-filled solid lines). The data presented in orange correspond to outflow measurements with a minimum outflow velocity of $150\, {\rm km\, s^{-1}}$, while the data presented in blue correspond to outflow measurements with a minimum outflow velocity of $50\, {\rm km\, s^{-1}}$. In our calculations, hatched shaded regions correspond to the bootstrap-generated $90\%$ confidence interval on the medians at a given mass. For both velocity thresholds, our measurements agree very well, to within $\approx0.1\,{\rm dex}$. }
\label{fig:apdx:literature_nelson19}
\end{figure}

 Figure \ref{fig:apdx:literature_nelson19} shows a comparison of $z\approx2$ $10{\rm kpc}$-scale mass loading factors as a function of $M_{\star}$ in \tng\ for two different choices of outflow velocity cut: $50\, {\rm km\,s^{-1}}$, and $150\, {\rm km\,s^{-1}}$. Utilising a very similar method, our measurements agree very well, to within $\approx0.1\,{\rm dex}$ for both velocity thresholds, across the range of stellar masses shown. As discussed in \S\ref{sec:methods:flows}, in the main body of this paper we do not enforce a minimum velocity threshold for the outflows. With experimentation, we find that imposing a velocity threshold does not qualitatively influence our results.

\section{Galaxy density profiles}\label{sec:apdx:densityprofs}
\begin{figure}
\includegraphics[width=0.95\columnwidth]{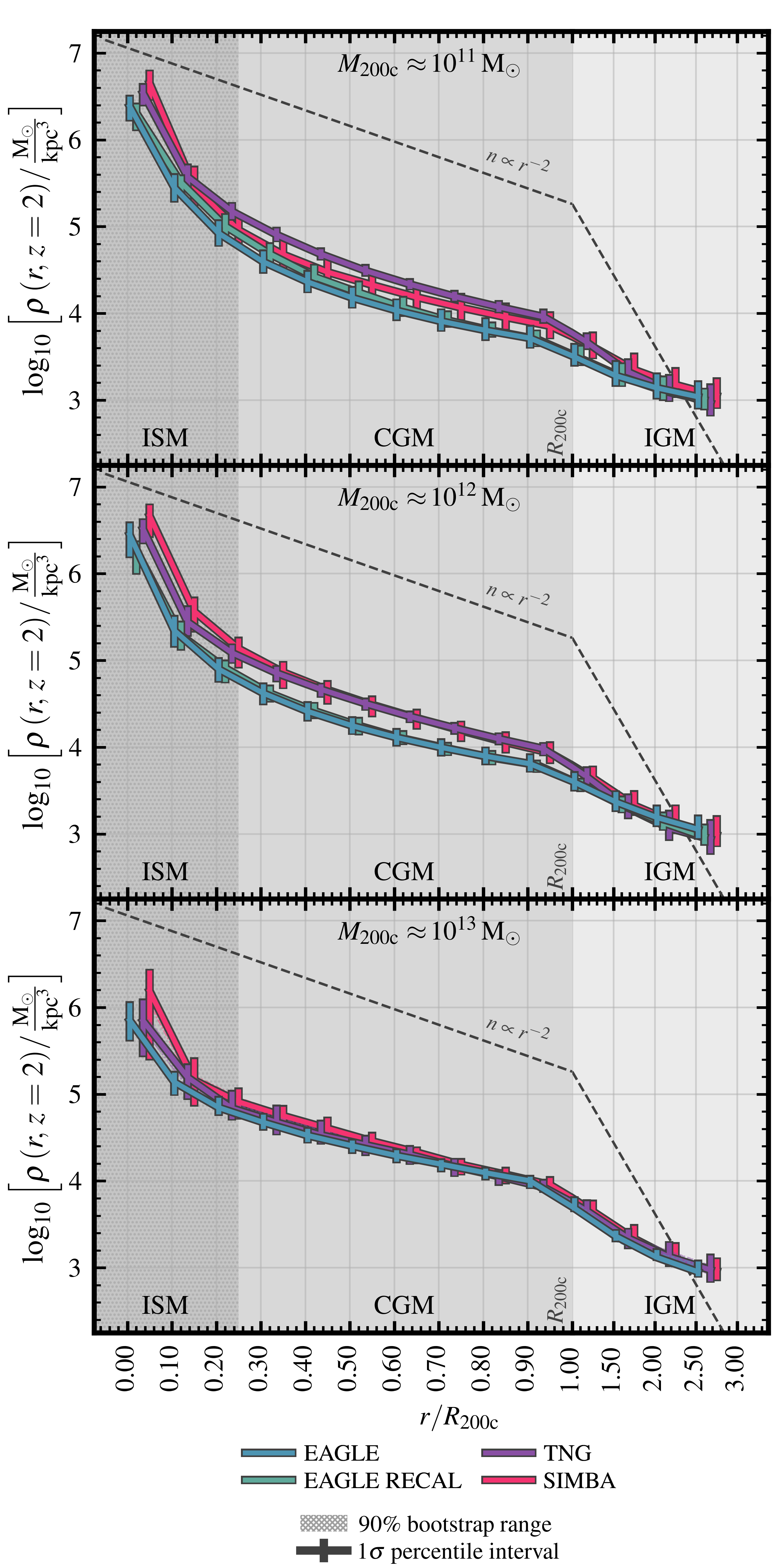}
\caption[]{Radial density profiles of galaxies at $z=2$ in \eagle\ (blue), \eagle-RECAL (teal), \tng\ (purple) and \simba\ (pink).Each column represents a different bin in halo mass, increasing top to bottom. Error-bars correspond to the $16^{\rm th} - 84^{\rm th}$ percentile range in inflow rates at a given mass, and hatched regions correspond to the bootstrap-generated $90\%$ confidence interval on the medians at a given mass. Gas density profiles are relatively similar between the simulations at this redshift, corresponding to the universally higher halo baryon fractions presented in Fig. \ref{fig:apdx:m200_scalings_z2}. }
\label{fig:apdx:r200_densityprof_z2}
\end{figure}

\begin{figure}
\includegraphics[width=0.95\columnwidth]{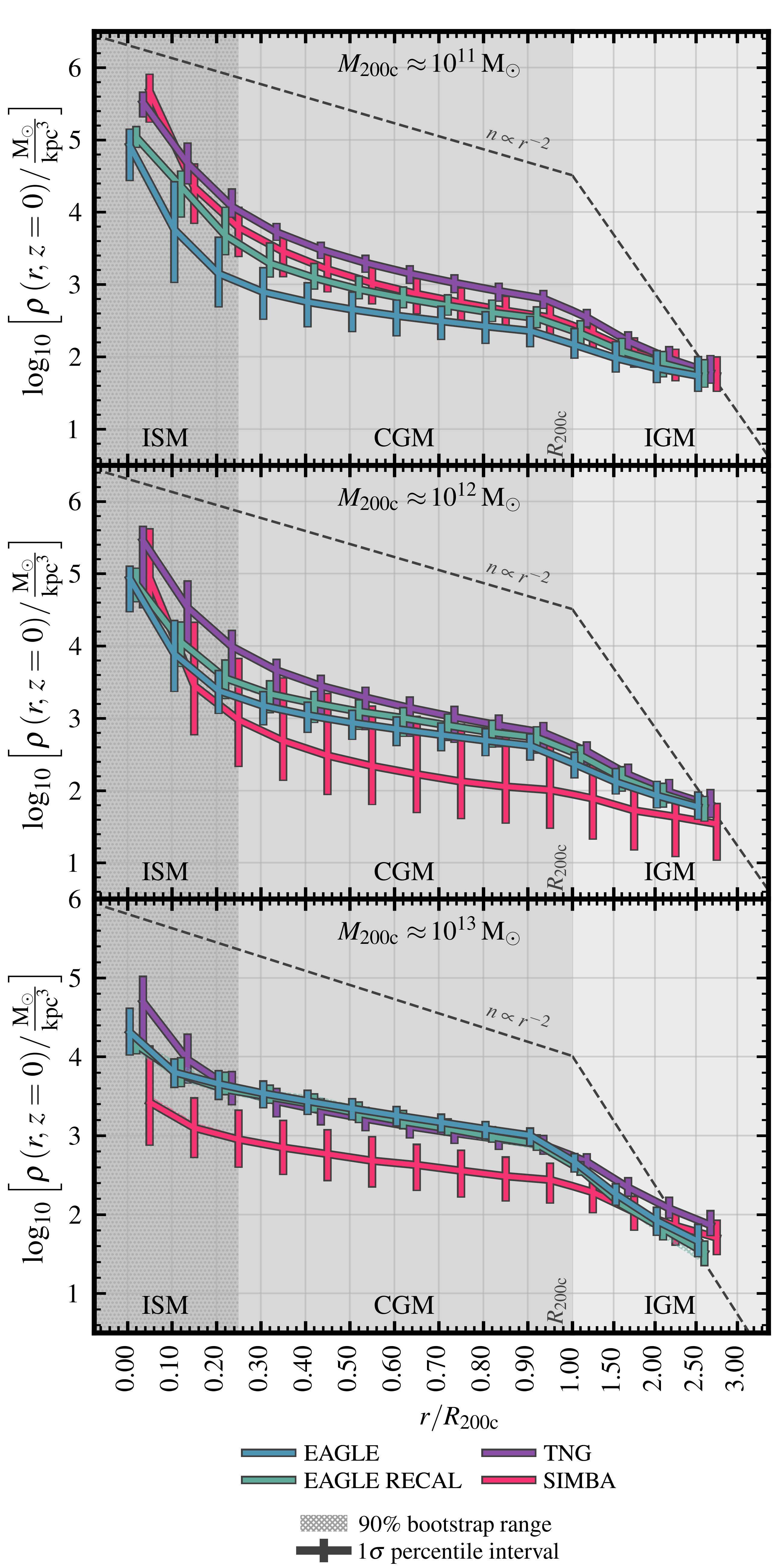}
\caption[]{Radial density profiles of galaxies at $z=0$ in \eagle\ (blue), \eagle-RECAL (teal), \tng\ (purple) and \simba\ (pink). Each row represents a different bin in halo mass, increasing top to bottom. Error-bars correspond to the $16^{\rm th} - 84^{\rm th}$ percentile range in inflow rates at a given mass, and hatched regions correspond to the bootstrap-generated $90\%$ confidence interval on the medians at a given mass. Density profiles diverge between the simulations at this redshift, with particularly low densities in \eagle\ in the lowest halo mass bin, and \simba\ for the two higher halo mass bins.}
\label{fig:apdx:r200_densityprof_z0}
\end{figure}

In Fig. \ref{fig:apdx:r200_densityprof_z2} and \ref{fig:apdx:r200_densityprof_z0}, we show the density profiles of galaxies in each of the simulations at $z=2$ and $z=0$ respectively. Each row represents a different bin in halo mass -- top: \logmhalo$\in[10.75,11.25]$; middle: \logmhalo$\in[11.75,12.25]$; bottom: \logmhalo$\in[12.75,13.25]$). Radial $x$-values are quoted as fractions of $R_{\rm 200c}$, with a broken scale above $1\times R_{\rm 200c}$ to include radial bins up to $3\times R_{\rm 200c}$. For reference, in each panel we include a dashed line indicating the expected outer slope of an isothermal $\beta $ profile with $\beta=2/3$; i.e. $\rho(r)\propto r^{-2}$. 

In general, density profiles at $z=2$ agree fairly well between the simulations. This corresponds to the universally higher baryon content of galaxies and their CGM at this redshift, as demonstrated in Fig. \ref{fig:apdx:m200_scalings_z2}. Considering the lowest mass bin at $z=0$, we note that the density of the CGM in \eagle\ is quite low -- $\approx1\,{\rm dex}$ lower than the predictions from \tng\ and \simba\ at the edge of the ISM. This low density allows for the entrainment of gas with outflows in \eagle, with the outflows being overpressurised relative to the surrounding medium. We observe a similar effect in \simba\ in the two higher mass bins, with CGM densities in the intermediate halo mass bin being $\approx1\,{\rm dex}$ lower than the predictions from \eagle\ and \tng. This signifies the efficient evacuation of the CGM via AGN feedback, which also allows subsequent outflows to be overpressurised relative to the CGM and entrain gas in the outflows towards larger scales. 

\section{Galaxy temperature profiles}\label{sec:apdx:tempprofs}
\begin{figure}
\includegraphics[width=0.95\columnwidth]{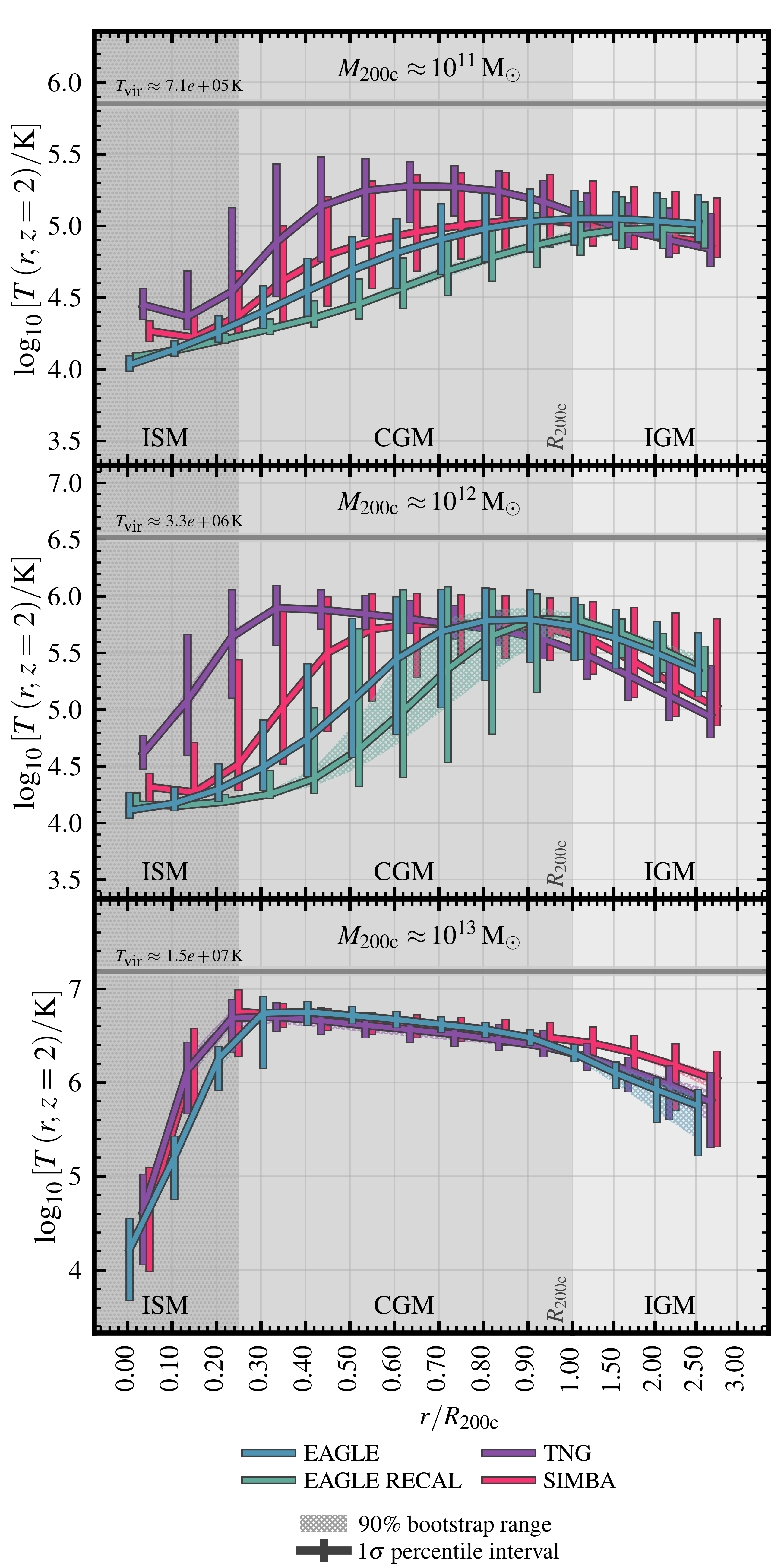}
\caption[]{Radial temperature profiles of galaxies at $z=$ in \eagle\ (blue), \eagle-RECAL (teal), \tng\ (purple) and \simba\ (pink). Each row represents a different bin in halo mass, increasing top to bottom. Error-bars correspond to the $16^{\rm th} - 84^{\rm th}$ percentile range in inflow rates at a given mass, and hatched regions correspond to the bootstrap-generated $90\%$ confidence interval on the medians at a given mass. Temperature predictions diverge between the simulations at this redshift in the two lower mass bins, particularly in the inner CGM.} 

\label{fig:apdx:r200_Tprof_z2}
\end{figure}

\begin{figure}
\includegraphics[width=0.95\columnwidth]{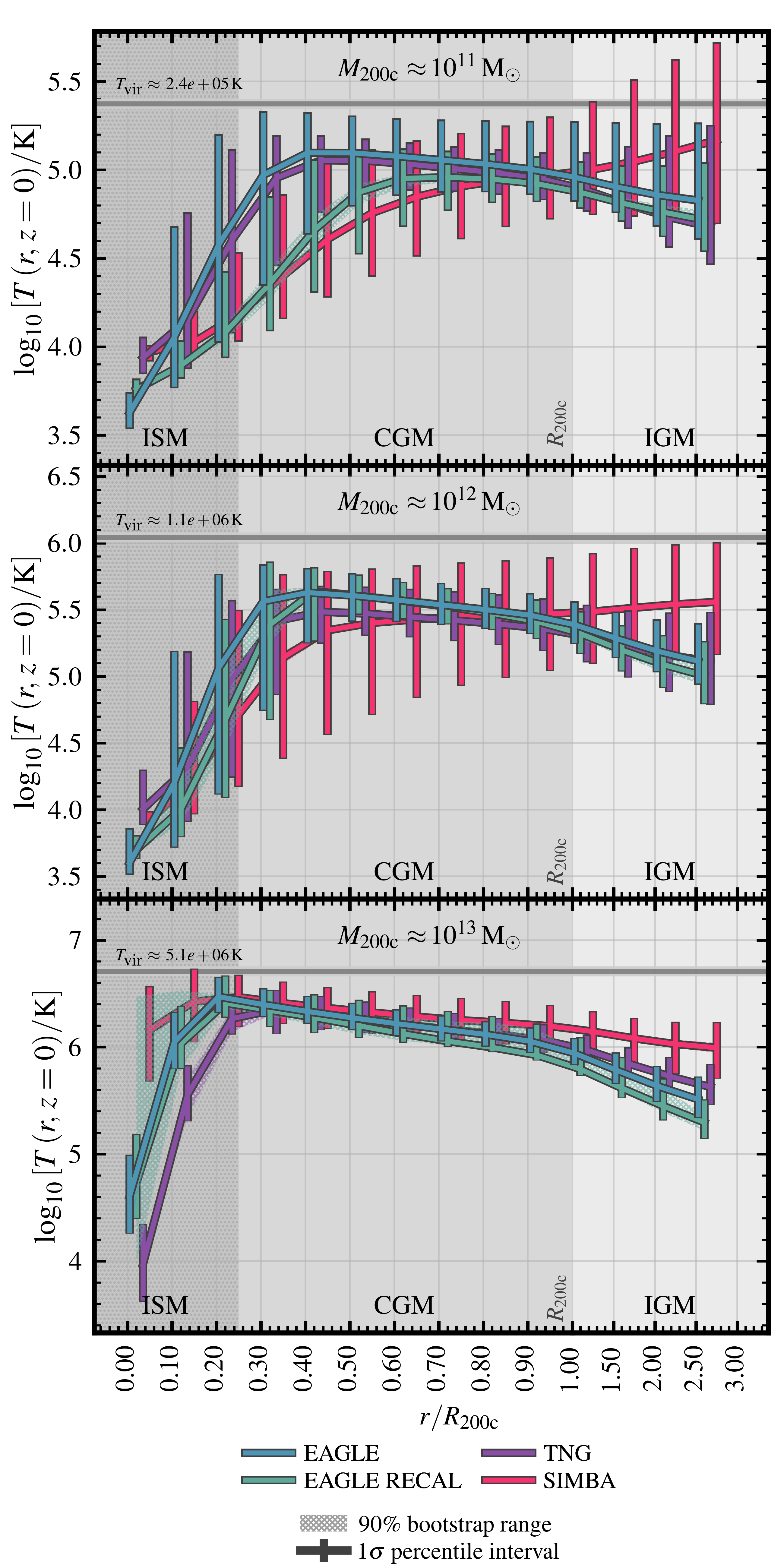}
\caption[]{Radial temperature profiles of galaxies at $z=0$ in \eagle\ (blue), \eagle-RECAL (teal), \tng\ (purple) and \simba\ (pink). Each row represents a different bin in halo mass, increasing top to bottom. Error-bars correspond to the $16^{\rm th} - 84^{\rm th}$ percentile range in inflow rates at a given mass, and hatched regions correspond to the bootstrap-generated $90\%$ confidence interval on the medians at a given mass. Temperature profiles are in slightly better agreement between the simulations compared to $z\approx2$, with generally hotter CGM. }
\label{fig:apdx:r200_Tprof_z0}
\end{figure}

\begin{figure*}
\includegraphics[width=\textwidth]{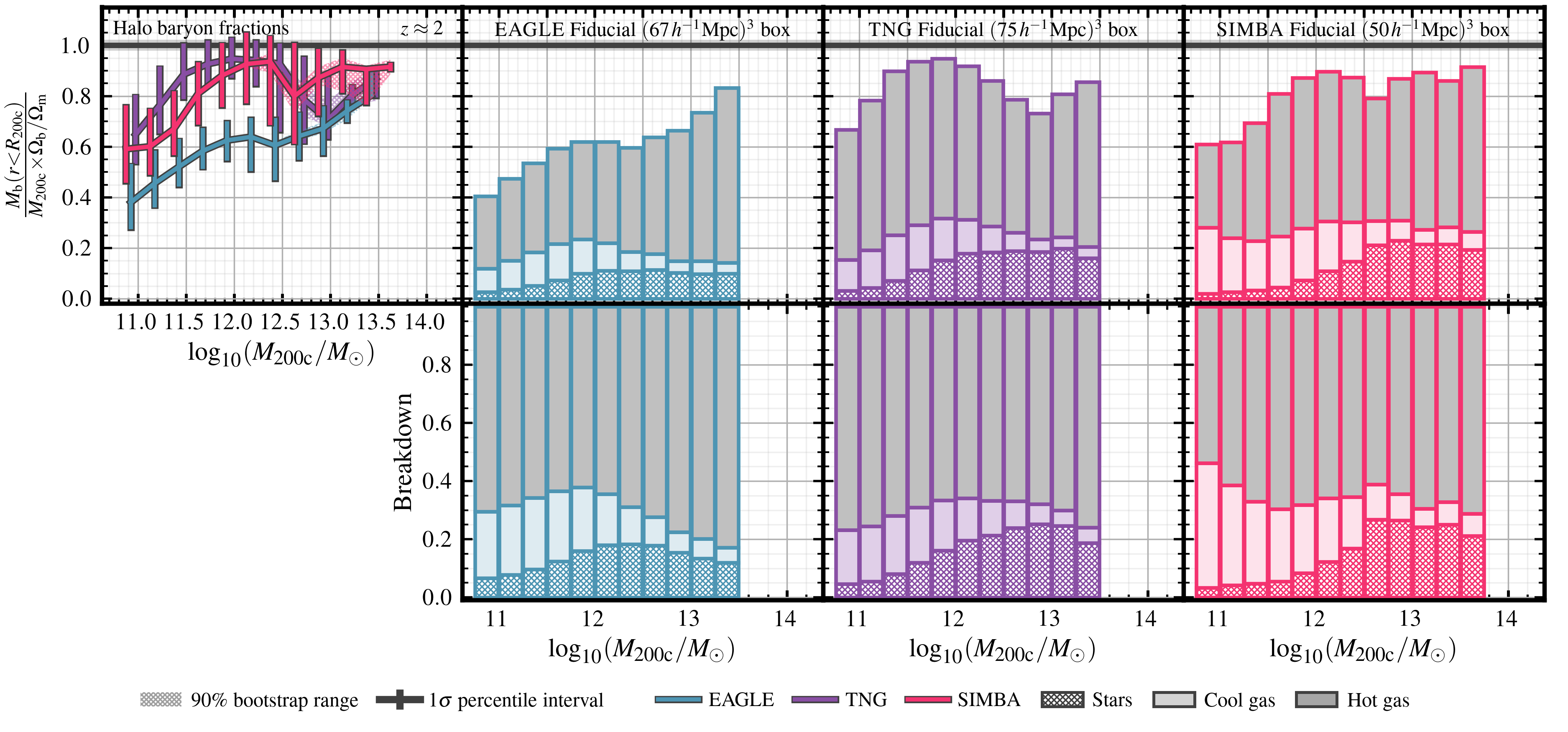}
\caption[]{The baryon content of haloes at $z\approx2$ in \eagle\ (blue), \tng\ (purple) and \simba\ (pink). {\it Top panel, $1^{\rm st}$ column}: The median total baryon fraction within $R_{\rm 200c}$ (stars, gas, black holes) of haloes as a function of $M_{\rm 200c}$ mass. {\it Top panels,  $2^{\rm nd}-4^{\rm th}$ columns}: Total baryon content for each simulation within $R_{\rm 200c}$ broken down into stars (hatched bars), ``cold'' gas (coloured shaded bars), and ``hot gas'' (grey bars) in each simulation. In the top panels, this baryon content is normalised by $f_{\rm b}\times M_{\rm 200c}$. {\it Bottom panels}: The bottom row of panels shows the same breakdown, but re-normalised by the actual baryonic mass in each halo, $M_{\rm bar} (r<R_{\rm 200c})$.  In each of the simulations in the left-hand panel, error-bars represent the $16^{\rm th}-84{\rm th}$ percentiles in halo baryon fractions for a given mass bin; the hatched regions represent the $90\%$ confidence interval on the median from 100 bootstrap re-samplings in each bin. The difference between the simulations in terms of $z\approx2$ baryon fractions are qualitatively similar compared to that shown at $z\approx0$ in Fig. \ref{fig:results1:m200_scalings_z0}, but the baryon content of haloes is universally higher. }
\label{fig:apdx:m200_scalings_z2}
\end{figure*}

In Fig. \ref{fig:apdx:r200_Tprof_z2} and \ref{fig:apdx:r200_Tprof_z0}, we show the temperature profiles of galaxies in each of the simulations at $z=2$ and $z=0$ respectively. Each row represents a different bin in halo mass -- top: \logmhalo$\in[10.75,11.25]$; middle: \logmhalo$\in[11.75,12.25]$; bottom: \logmhalo$\in[12.75,13.25]$). Radial $x$-values are quoted as fractions of $R_{\rm 200c}$, with a broken scale above $1\times R_{\rm 200c}$ to include radial bins up to $3\times R_{\rm 200c}$. For reference, in each panel we include a horizontal line indicating the expected virial temperature of haloes in this mass range, with $T_{\rm vir}\approx1.1\times 10^{6}\, {\rm K}\, \left(\mu/0.59\right) \left({M_{\rm halo}}/{10^{12}{\rm M}_{\odot}}\right)^{2/3} (1+z)$, from \citet{VandeVoort2017_chapter}.

As discussed in \S\ref{sec:results:2:outflows}, considering haloes with mass $M_{\rm 200c}\lesssim 10^{13}{\rm M}_{\odot}$ at $z\approx2$,  predictions for temperatures in the inner CGM are quite different between the simulations. In particular, in \eagle\ (and to some extent, \simba) there remains a significant presence of cold gas between $0.3-0.6\times R_{\rm 200c}$. Comparatively, in \tng, temperatures in this region typically exceed $10^{5}\, {\rm K}$. At lower redshift, the predictions between the simulations are relatively similar in the two lower mass bins (signifying a universal shift to a hotter CGM, and the dominance of hot-mode accretion).

\section{Halo baryon fractions at redshift 2}\label{sec:apdx:fbz2}

In this section, we add to the $z\approx0$ results shown in \S\ref{sec:results:1} to compare the total baryon content of haloes at as a function of halo mass, and how this baryon content is distributed between stars, cold gas, and hot gas in each of the simulations. As in the main body of this paper, in each simulation, we define ``cold gas'' as the gas that is either considered star-forming, or below $5\times10^{4} {\rm K}$, and ``hot gas'' as the remaining gas within $R_{\rm 200c}$ that does not meet this criteria. 

At this redshift, the baryon content of haloes is universally higher in the simulations -- for $M_{\rm 200c}\gtrsim10^{11.5}{\rm M}_{\odot}$, all simulations predict a baryon fraction above $50\%$ of the cosmic mean. Similar to $z\approx0$, the baryon content of \eagle galaxies at low halo mass -- $M_{\rm 200c}\lesssim10^{12}{\rm M}_{\odot}$ -- is lower than the predictions from \simba\ and \tng, indicating that the ejective and preventative influence of stellar feedback is already at play by $z\approx2$. The transition masses \Mt{1} and \Mt{2} are similar to those reported at $z\approx0$, with the exception of \simba, where the  \Mt{1} mass at $z\approx2$ increases to $\approx10^{12}{\rm M}_{\odot}$ (as opposed to $\approx10^{11.3}{\rm M}_{\odot}$ at $z\approx0$, with \Mt{2} remaining at $\approx10^{12.8}{\rm M}_{\odot}$). For \simba, this indicates that AGN feedback becomes more efficient at lower masses towards $z\approx0$. 

\bsp	
\label{lastpage}
\end{document}